\documentclass[fleqn,12pt]{article}
\usepackage{amssymb,epsfig}

\def\beq   {\begin{equation}}
\def\eeq   {\end{equation}}
\def\beqd  {\begin{displaymath}}
\def\eeqd  {\end{displaymath}}
\def\beqaa {\begin{eqnarray}}
\def\eeqaa {\end{eqnarray}}

\begin{document}
%------------------------------------------------------------------------
\pagestyle{empty}

\begin{flushright}
  CERN-PH-TH/2006-130\\
  DCPT-06-98\\
  IPPP-06-49\\
  UWThPh-2006-17 \\
  hep-ph/0608065
\end{flushright}

\vfill

\begin{center}

\textbf{\Large
CP asymmetries in chargino production and decay:
the three-body decay case
}\\

\vspace{10mm}

{\large
A.~Bartl$^1$, H.~Fraas$^2$, S.~Hesselbach$^2$,
K.~Hohenwarter-Sodek$^1$, T.~Kernreiter$^1$, G.~Moortgat-Pick$^3$
}

\vspace{6mm}

{\itshape
$^1$ Institut f\"ur Theoretische Physik, Universit\"at Wien, A-1090
Vienna, Austria\\
$^2$ Institut f\"ur Theoretische Physik und Astrophysik,
Universit\"at W\"urzburg,\\
{\it Am Hubland, D-97074 W\"urzburg, Germany}\\
$^{3}$~TH Division, Physics Department, CERN, CH-1211 Geneva 23,
Switzerland
}

\end{center}

\vfill

\begin{abstract}

We study CP violation in chargino production and decay 
in the Minimal Supersymmetric Standard Model (MSSM)
with complex parameters
at an $e^+e^-$ linear collider with
longitudinally polarized beams. We investigate CP-sensitive asymmetries
by means of triple product correlations
and study their dependence
on the complex parameters $M_1$ and $\mu$. We
give numerical predictions for the asymmetries and their
measurability at the future International Linear Collider.
Our results show that the CP asymmetries can be measured 
in a large region of the MSSM parameter space.

\end{abstract}

\vfill

\newpage
\pagestyle{plain}

%%%%%%%%%%%%%%%%%%%%%%%%%%%%%%%%%
\section{Introduction}
%%%%%%%%%%%%%%%%%%%%%%%%%%%%%%%%%

In the Minimal Supersymmetric Standard Model (MSSM) \cite{Haber:1984rc} the
supersymmetric partners of the gauge bosons and Higgs bosons
with the same electric charge mix and form the neutralinos 
$\tilde\chi^0_i$ ($i=1,\dots,4$) and the 
charginos $\tilde\chi^+_k$ ($k=1,2$), as the neutral and charged 
mass eigenstates, respectively.
The charginos and the neutralinos are of particular
interest, as they will presumably be among the lightest supersymmetric (SUSY) particles.
One of the main goals of the International Linear Collider (ILC)
\cite{LC,Aguilar-Saavedra:2005pw} will be the determination of the underlying SUSY
parameters.
Those parameters that enter the neutralino/chargino
system at tree level are the gaugino mass parameters $M_1$ and $M_2$, the
higgsino mass parameter $\mu$, and the ratio of the vacuum 
expectation values of the Higgs fields, $\tan\beta$. 
Among these parameters
$M_1$ and $\mu$ can be complex, while $M_2$ and $\tan\beta$ can be
chosen real. In \cite{ref13,ref14} 
methods have been developed to
determine the parameters in the neutralino and chargino system with
and without CP violation by measurements of the 
neutralino and chargino masses and their production cross
sections.

The phases $\phi_\mu$ and $\phi_{M_1}$ of  $\mu$ and $M_1$
may be constrained or correlated by the experimental upper bounds on
the electric dipole moments (EDMs).
These constraints, however, are rather model-dependent~\cite{Ibrahim:2001yv}. 
While the restriction on the phase $\phi_\mu$,
due to the electron EDM, is rather severe in a constrained MSSM with selectron
masses of the order of $100$~GeV \cite{edm}, 
it may disappear if lepton-flavour-violating terms
in the MSSM Lagrangian are included \cite{Bartl:2003ju}.
Recently it has been pointed out that for large trilinear scalar couplings $A$ 
we can simultaneously fulfil the EDM constraints of electron, neutron, and 
of the atoms $^{199}$Hg and $^{205}$Tl where, at the same time, 
$\phi_\mu\sim O(1)$ \cite{YaserAyazi:2006zw}.
The size of the phase $\phi_{M_1}$, on the other hand, is less strongly restricted
in the MSSM.
Thus, the CP phases $\phi_{M_1}$ and $\phi_\mu$
can have a big influence on the production and decay of charginos
and neutralinos at the ILC. In particular, they give rise
to CP-sensitive observables that may be accessible
at future collider experiments. Measurements of CP-sensitive 
observables are necessary to prove that CP is violated. Furthermore,
only the inclusion of CP-sensitive observables allows us to deduce the 
underlying model parameters in an 
unambiguous way. In neutralino production with subsequent decay, CP-sensitive 
observables based on triple product correlations 
have been investigated in 
\cite{Kizukuri:1990iy,Aguilar-Saavedra:2004dz,Bartl:2004jj,Bartl:2005uh}. 
Also for the case of chargino production and decay, 
various CP-sensitive observables have been studied. Decay rate
asymmetries in chargino decays have been studied in \cite{Yang:2002am}
and CP asymmetries in decay chains of sneutrinos involving charginos in
\cite{Aguilar-Saavedra:2004ru}.
CP-sensitive asymmetries based on triple product correlations
have been analysed for the subsequent two-body decays
$\tilde{\chi}^-_j \to \tilde{\chi}^0_1 W^-$ \cite{Kittel:2004kd} and 
$\tilde{\chi}^-_j \to \tilde{\nu}_\ell \ell^-$ \cite{Bartl:2004vi}.
For the case of transverse $e^\pm$ beam polarization
azimuthal asymmetries have been studied for the same
two-body decays, showing a pronounced dependence on 
$\phi_{M_1}$ and $\phi_\mu$ \cite{Bartl:2004xy}.
In the present paper we extend previous 
investigations of CP violation in chargino 
production and decay to the case of chargino three-body decays. 

We study the production processes
\begin{equation}\label{process1}
e^+e^-\rightarrow \tilde{\chi}^+_1\tilde{\chi}^-_k, \qquad
k=1,2~,
\end{equation}
at a linear collider with longitudinal beam polarizations, and
subsequent leptonic or hadronic three-body decays of the 
$\tilde{\chi}^+_1$,
\begin{equation}\label{process2}
\tilde{\chi}^+_1 \to \tilde{\chi}^0_1~\nu~\ell^+~,\quad \ell=e,\mu~,
\end{equation}
and
\begin{equation}\label{process3}
\tilde{\chi}^+_1 \to \tilde{\chi}^0_1~\bar{s}~c~,
\end{equation}
where we assume that the momenta $\vec{p}_{\tilde\chi_1^+}$, 
$\vec{p}_{\ell}$, $\vec{p}_{c}$ and $\vec{p}_{s}$ of the 
associated particles can be measured or reconstructed.
We study two T-odd observables based on
triple product correlations of momentum vectors:
\begin{equation} \label{tpdeflep}
\mathcal{T}_\ell=\vec{p}_{\ell^+}
 \cdot (\vec{p}_{e^-} \times \vec{p}_{\tilde{\chi}^+_1})~,
\end{equation}
\begin{equation} \label{tpdefhad}
\mathcal{T}_q=\vec{p}_{\bar{s}}\cdot (\vec{p}_{c}\times\vec{p}_{e^-})~.
\end{equation}
The triple product $\mathcal{T}_\ell$,
Eq.~(\ref{tpdeflep}), relates momenta of initial,
intermediate and final particles, whereas $\mathcal{T}_q$,
Eq.~(\ref{tpdefhad}), uses
only momenta from the initial and final states. Therefore, 
both triple products depend in a different way on the 
production and decay processes.

The triple product $\mathcal{T}_\ell$, Eq.~(\ref{tpdeflep}), involves the 
momentum of the
decay lepton that usually can be very accurately 
measured. However, the momentum of the chargino has to be reconstructed
with information from the 
decay of the second chargino $\tilde\chi^-_k$ \cite{Bartl:2005uh}.
For the triple product
$\mathcal{T}_q$, Eq.~(\ref{tpdefhad}), it is necessary to identify the $c$-quark, which 
is expected to be possible with reasonable efficiency and purity \cite{TDR,XellaHansen:2003sw,Damerell:1996sv}.
To derive the CP-violating asymmetry also the charge of the $c$-quark has to be detected, 
which can be done with specific vertex detectors \cite{TDR,Damerell,Abe:2001dr}.
The corresponding T-odd asymmetries are defined by
\begin{equation} \label{eq:Todd}
A_T(\mathcal{T}_{\ell,q})=
\frac{N[\mathcal{T}_{\ell,q}> 0]-N[\mathcal{T}_{\ell,q} < 0]}
{N[\mathcal{T}_{\ell,q}> 0]+N[\mathcal{T}_{\ell,q} < 0]}~,
\end{equation}
where $N[\mathcal{T}_{\ell,q} > (<)~ 0]$ is the number
of events for which $\mathcal{T}_{\ell,q}> (<) ~0$.

Finally we recall that a non-zero value of the 
T-odd asymmetries does not 
immediately imply that the CP symmetry is violated since final-state 
interactions give rise (although only at the one-loop level) to the same asymmetries.
However, a genuine signal of CP violation can be obtained when 
one combines $A_T(\mathcal{T}_{\ell,q})$ with the corresponding asymmetry
$\bar{A}_T(\mathcal{T}_{\ell,q})$ for the charge-conjugated processes.
Then in the CP asymmetries
\begin{equation}\label{ACP}
A_{\rm CP}(\mathcal{T}_{\ell,q})=
\frac{A_T(\mathcal{T}_{\ell,q})-\bar{A}_T(\mathcal{T}_{\ell,q})}{2}~,
\end{equation}
the effect of final-state interactions cancels out. 

The paper is organized as follows.
In Section 2 we briefly
recall the formalism, which we use to calculate
the cross sections and the CP asymmetries.
We present
our numerical results in Section 3. Finally, we summarize 
and conclude in Section 4.

%%%%%%%%%%%%%%%%%%%%%%%%%%%%%%%%%%%%%%%%%%%%%%%%%%%%%%%%
\section{Cross section and CP asymmetries}
%%%%%%%%%%%%%%%%%%%%%%%%%%%%%%%%%%%%%%%%%%%%%%%%%%%%%%%%

\begin{figure}[htb]
\centering
\epsfig{file=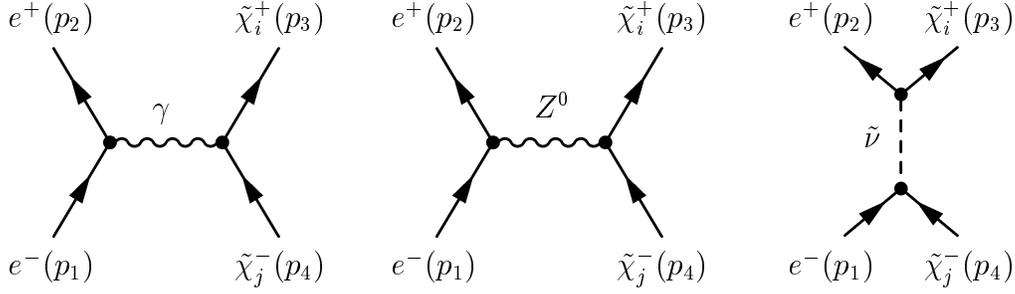}
\caption{\label{Fig:FeynProd}Feynman diagrams of the production process
  $e^{+}e^{-}\to\tilde{\chi}^+_i\tilde{\chi}^-_j$.}
\end{figure}

\begin{figure}[htb]
\centering
\epsfig{file=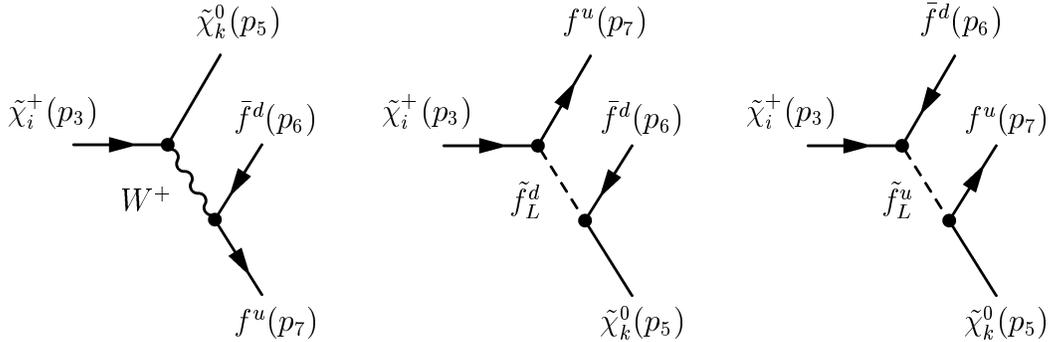}
\caption{\label{Fig:FeynDecay}Feynman diagrams of the three-body decay
$\tilde{\chi}^+_i\to \tilde{\chi}^0_k \bar{f^d} f^u$, where 
$f^d=e, \mu, s$ and $f^u=\nu_e,\nu_\mu,c$.}
\end{figure}

The chargino production process (\ref{process1}) proceeds
via $\gamma$ and $Z^0$ exchange in the $s$-channel
and via $\tilde{\nu}_e$ exchange in the $t$-channel
(Fig.~\ref{Fig:FeynProd}).
The decay processes (\ref{process2}) and (\ref{process3})
 contain contributions 
from $W^+$, $\tilde{\ell}_L$ $(\ell=e,\mu)$ and
$\tilde{\nu_\ell}$ exchange in the leptonic case
and from $W^+$, $\tilde{c}_L$  and $\tilde{s}_L$ exchange 
in the hadronic case (Fig.~\ref{Fig:FeynDecay}). 
The interaction Lagrangians for these processes 
can be found, for instance, in \cite{Moortgat-Pick:1998sk}.


%%%%%%%%%%%%%%%%%%%%%%%%%%%%%%%%%%%%%%%%%%
\subsection{Cross section}
%%%%%%%%%%%%%%%%%%%%%%%%%%%%%%%%%%%%%%%%%%

For the calculation of the squared amplitude of the whole process 
$e^+e^-\to \tilde{\chi}^+_i\tilde{\chi}^-_j\to
\tilde{\chi}^0_1 \bar{f}^d f^u \tilde{\chi}^-_j$, we use the 
spin-density matrix formalism \cite{Moortgat-Pick:1998sk,Haber:1994pe}.
The squared amplitude can then be written as
\begin{equation}
|T|^2=2 |\Delta(\tilde\chi^+_i)|^2~
\sum_{\lambda_i,\lambda'_i}~
{\rho_{P}}^{\lambda_i\lambda'_i}~
{\rho_{D}}_{\lambda'_i\lambda_i}~,
\label{eq:matrixelement}
\end{equation}
with the propagator
$\Delta(\tilde{\chi}^+_i)=1/[p_{\tilde\chi_i^+}^2-m^2_i+im_i\Gamma_i]$.
Here, $\lambda_{i}$, $\lambda_{i}'$, 
$m_i$, $\Gamma_i$ denote the helicities, masses
and widths of the chargino $\tilde\chi_i^+$. The factor 2 in Eq.~(\ref{eq:matrixelement})
is due to the summation over the helicities of chargino $\tilde\chi^-_j$, whose
decay is not considered.
The squared amplitude is composed of the unnormalized 
spin-density matrices $\rho_P$ for the production and $\rho_D$ for the
decay, which carry the
helicity indices $\lambda_i,\lambda'_i$ of the chargino $\tilde{\chi}^{+}_i$.
Introducing a set of polarization basis 4-vectors
$s^a_{\chi_i}\;(a=1,2,3)$ for the charginos $\tilde\chi^+_i$, 
where $s^3_{\chi_i}$ describes the longitudinal
polarization and $s^1_{\chi_i}$, $s^2_{\chi_i}$ the
transverse polarization in and perpendicular to the production plane,
respectively, and
which fulfil the orthonormality relations 
$s^a_{\chi_i}\cdot s^b_{\chi_i}=-\delta^{ab}$ and
$s^a_{\chi_i}\cdot p_{\chi_i}=0$,
the density matrices can be 
expanded in terms of the Pauli matrices:
\begin{equation}
{\rho_{P}}^{\lambda_i\lambda'_i} = \delta_{\lambda_i\lambda'_i} P+
\sum^3_{a=1} \sigma^a_{\lambda_i\lambda'_i}\Sigma_P^a~,
\label{eq:rhoP}
\end{equation}
\begin{equation}
{\rho_{D}}_{\lambda'_i\lambda_i} = \delta_{\lambda'_i\lambda_i} D+
\sum^3_{a=1} \sigma^a_{\lambda'_i\lambda_i}\Sigma_D^a~.
\label{eq:rhoD}
\end{equation}
Then the squared amplitude is given by
\begin{eqnarray}
|T|^2 = 4|\Delta(\tilde{\chi}^+_i)|^2
 \left\{ P(\tilde{\chi}^+_i\tilde{\chi}^-_j) D(\tilde{\chi}^+_i)
             +
            \sum^3_{a=1}\Sigma^a_P(\tilde{\chi}^+_i)
                    \Sigma^a_D(\tilde{\chi}^+_i)\right\},
\label{Tsquared}
\end{eqnarray}
where $P(\tilde{\chi}^+_i\tilde{\chi}^-_j)$ and $D(\tilde{\chi}^+_i)$
are those parts of the spin density production and decay matrices,
that are independent of the polarization of the charginos.
The contributions $\Sigma^a_P(\tilde{\chi}^+_i)$ and
$\Sigma^a_D(\tilde{\chi}^+_i)$ depend on the polarization
vector $s^a$ of the decaying chargino $\tilde{\chi}^+_i$.
The full expressions for the quantities 
$P(\tilde{\chi}^+_i\tilde{\chi}^-_j)$, $D(\tilde{\chi}^+_i)$,
$\Sigma^a_P(\tilde{\chi}^+_i)$ and
$\Sigma^a_D(\tilde{\chi}^+_i)$ can be found in
\cite{Moortgat-Pick:1998sk}.
Finally, the differential cross section is given by
\begin{equation}\label{crosssection}
d\sigma =
 \frac{1}{8E^2_b}|T|^2
 (2\pi)^4 \delta^4\left(p_1 + p_2 - \sum_{i=4}^7 p_i\right)d
  \mbox{lips}(p_3 \cdots p_{7})~,
\end{equation}
where $E_b$ is the beam energy and $d\mbox{lips}(p_3 \cdots p_{7})$ 
is the Lorentz-invariant phase-space element.

%%%%%%%%%%%%%%%%%%%%%%%%%%%%%%%%%%%%%%%
\subsection{CP asymmetries}
\label{subasy}
%%%%%%%%%%%%%%%%%%%%%%%%%%%%%%%%%%%%%%%

The T-odd asymmetries defined in Eq.~(\ref{eq:Todd})
are calculated as
\begin{equation} \label{Asy}
A_T(\mathcal{T}_{\ell,q}) = \frac{\int\mathrm{sign}\{ \mathcal{T}_{\ell,q} \}
~|T|^2~d\mbox{lips}}{{\int}~|T|^2~d\mbox{lips}}~,
\end{equation}
where we weight the sign of the triple product
correlations in Eqs.~(\ref{tpdeflep}) and (\ref{tpdefhad}) with
the associated squared amplitude. 
Since in the numerator 
$\int\mathrm{sign}\{ \mathcal{T}_{\ell,q} \}P(\tilde{\chi}^+_i\tilde{\chi}^-_j) 
D(\tilde{\chi}^+_i) d\mbox{lips}=0$
and in the denominator
$\int\Sigma^a_P(\tilde{\chi}^+_i)\Sigma^a_D(\tilde{\chi}^+_i)d\mbox{lips}=0$,
we obtain by inserting the squared amplitude, Eq.~(\ref{Tsquared}), into
Eq.~(\ref{Asy}):
\begin{equation}\label{Asyparts}
A_T(\mathcal{T}_{\ell,q})=\frac{\int\mathrm{sign}\{ \mathcal{T}_{\ell,q} \}
\Sigma^a_P(\tilde{\chi}^+_i)\Sigma^a_D(\tilde{\chi}^+_i) 
d\mbox{lips}}{{\int} P(\tilde{\chi}^+_i\tilde{\chi}^-_j) 
D(\tilde{\chi}^+_i) d\mbox{lips}}~.
\end{equation}
We split
$\Sigma^a_P(\tilde{\chi}^+_i)$ and $\Sigma^a_D(\tilde{\chi}^+_i)$
into the T-odd terms
$\Sigma^{a,\mathrm{O}}_{P}(\tilde{\chi}^+_i)$ and
$\Sigma^{a,\mathrm{O}}_{D}(\tilde{\chi}^+_i)$, which contain the
respective triple product, and T-even terms
$\Sigma^{a,\mathrm{E}}_{P}(\tilde{\chi}^+_i)$ and
$\Sigma^{a,\mathrm{E}}_{D}(\tilde{\chi}^+_i)$
without triple products:
\begin{equation} \label{sigmaO}
 \Sigma^a_P(\tilde{\chi}^+_i) =
  \Sigma^{a,\mathrm{O}}_P(\tilde{\chi}^+_i) +
  \Sigma^{a,\mathrm{E}}_P(\tilde{\chi}^+_i)~, \qquad
 \Sigma^a_D(\tilde{\chi}^+_i) =
  \Sigma^{a,\mathrm{O}}_D(\tilde{\chi}^+_i) +
  \Sigma^{a,\mathrm{E}}_D(\tilde{\chi}^+_i)~.
\end{equation}
The terms of $|T|^2$, Eq.~(\ref{Tsquared}), which
contribute to the numerator of $A_T$ are
\begin{equation} \label{CPinT}
\Sigma^{a,\mathrm{O}}_P(\tilde{\chi}^+_i)
                    \Sigma^{a,\mathrm{E}}_D(\tilde{\chi}^+_i)
   + \Sigma^{a,\mathrm{E}}_P(\tilde{\chi}^+_i)
       \Sigma^{a,\mathrm{O}}_D(\tilde{\chi}^+_i)~,
\end{equation}
where the first (second) term is sensitive to
the CP phases in the production (decay) process
of the chargino $\tilde{\chi}^+_i$.
The explicit expressions for the T-odd and T-even
contributions in Eq.~(\ref{CPinT}) 
are given in Appendix A. (The analytical expressions of the
quantities $P(\tilde{\chi}^+_i\tilde{\chi}^-_j)$ and 
$D(\tilde{\chi}^+_i)$ can be found in~\cite{Moortgat-Pick:1998sk}.)
With $A_T(\mathcal{T}_{\ell,q})$ we calculate the corresponding
CP asymmetries $A_{\rm CP}(\mathcal{T}_{\ell,q})$
according to Eq.~(\ref{ACP}).

%%%%%%%%%%%%%%%%%%%%%%%%%%%%%%%%%
\section{Numerical results}
%%%%%%%%%%%%%%%%%%%%%%%%%%%%%%%%%

In this section we give numerical results for the CP asymmetries 
$A_{\rm CP}(\mathcal{T}_{\ell,q})$, Eq.~(\ref{ACP}), for the reactions 
(\ref{process1}), (\ref{process2}), (\ref{process3}),
at an $e^+ e^-$ linear collider with centre-of-mass
energy $\sqrt{s}=500$~GeV and longitudinally polarized beams.
We analyse the hadronic decay 
$\tilde{\chi}^+_1 \to \tilde{\chi}^0_1 \bar{s} c$ and the leptonic
decays $\tilde{\chi}^+_1 \to \tilde{\chi}^0_1 \ell^+ \nu$, $\ell=e,\mu$.
To this end, we consider three scenarios (see Tables~\ref{ScenarioA},
\ref{ScenarioB} and \ref{ScenarioC}) for which 
$m_{\tilde{\chi}^+_1} < m_W + m_{\tilde{\chi}^0_1}$ and
$m_{\tilde{\chi}^+_1} < m_{\tilde{f}_L^{u,d}}$ to rule out two-body
decays of $\tilde{\chi}^+_1$.
The chargino decay widths and branching ratios have been calculated
with the computer program SPheno \cite{Porod:2003um}.

The statistical significance to which $A_{\rm CP}$ can be determined
to be non-zero can be estimated in the following way:
Assuming that the statistical errors of $A_T$ \cite{Desch:2006xp}
and $\bar{A}_T$ are independent of each other,
the errors of $A_T$ and $\bar{A}_T$ are added in quadrature.
The absolute error of $A_{\rm CP}$ is then given by 
\begin{equation}
\Delta A_{\rm CP} = \mathcal{N}_\sigma 
  \frac{\sqrt{1 - A_{\rm CP}^2}}{\sqrt{2 \sigma \mathcal{L}_\mathrm{int}}}~,
\end{equation}
where $\mathcal{N}_\sigma$ denotes the respective number of standard
deviations, 
$\sigma = \sigma(e^+e^- \to \tilde{\chi}^+_1\tilde{\chi}^-_j) \cdot
B(\tilde{\chi}^+_1 \to \tilde{\chi}^0_1 f' \bar{f})$
being the corresponding cross section of the
combined production and decay processes and $\mathcal{L}_\mathrm{int}$ is
the integrated luminosity, where we assume 
$\mathcal{L}_\mathrm{int}=500$~fb$^{-1}$ in the theoretical estimates below.
For $A_{\rm CP} \lesssim 10\%$, i.e.\ $A_{\rm CP}^2 \lesssim 0.01$, it is
$\Delta A_{\rm CP} = 
 \mathcal{N}_\sigma/\sqrt{2\sigma \mathcal{L}_\mathrm{int}}$
in good approximation. If we require $A_{\rm CP} > \Delta A_{\rm CP}$ for 
$A_{\rm CP}$ to be measurable we obtain
\begin{equation} \label{nsigma}
\mathcal{N}_\sigma = \sqrt{2 A_{\rm CP}^2 \sigma \mathcal{L}_\mathrm{int}}~.
\end{equation}

\begin{table}[t]
\begin{center}
\begin{tabular}{|c|c|} \hline
\multicolumn{2}{|c|}{scenario A}\\
\hline\hline
$M_2$             & 280 \\ \hline
$|\mu|$           & 200 \\ \hline
$\tan\beta$       & 5   \\ \hline
$m_{\tilde{\nu}}$ & 250 \\ \hline
$m_{\tilde{u}_L}$ & 500 \\ \hline
\end{tabular}
\begin{tabular}{|c|c||c|c|c|c|c|c|} \hline
$\phi_{\mu}$ & $\phi_{M_1}$ &
$m_{\tilde{\chi}^0_1}$ & $m_{\tilde{\chi}^0_2}$ & $m_{\tilde{\chi}^0_3}$ 
& $m_{\tilde{\chi}^0_4}$  
&
$m_{\tilde{\chi}^\pm_1}$ & $m_{\tilde{\chi}^\pm_2}$\\ \hline\hline
0 &            0    & 119.3 & 184.3 & 205.9 & 322.7 & 166.2 & 322.1 \\
0 & $\frac{\pi}{2}$ & 126.3 & 176.0 & 210.0 & 323.0 & 166.2 & 322.1 \\
0 & $ \pi   $       & 135.3 & 166.7 & 213.0 & 321.3 & 166.2 & 322.1 \\ \hline
$\frac{\pi}{2}$ &           0     & 127.6 & 187.4 & 208.5 & 316.0 & 177.4 &
316.0  \\
$\frac{\pi}{2}$ & $\frac{\pi}{2}$ & 134.8 & 178.3  & 213.0 & 315.3 & 177.4 &
316.0  \\
$\frac{\pi}{2}$ & $\pi$           & 129.6 & 176.7 & 217.8 & 315.0 & 177.4  &
316.0  \\\hline
$\pi$ &       0         & 134.6 & 190.4 & 212.9 & 308.2 & 189.2  & 309.1 \\
$\pi$ & $\frac{\pi}{2}$ & 130.3 & 186.0 & 219.9 & 307.9 & 189.2  & 309.1 \\
$\pi$ & $\pi$           & 126.0 & 183.3 & 225.1 & 307.5 & 189.2  & 309.1  \\
\hline
\end{tabular}
\end{center}
\caption{Input parameters $M_2$, $|\mu|$, $\tan\beta$,
$m_{\tilde{\nu}}$ and $m_{\tilde{u}_L}=m_{\tilde{c}_L}$
and neutralino and chargino masses for scenario A
for different values of $\phi_\mu$ and 
$\phi_{M_1}$. 
$|M_1|$ is fixed by the GUT-inspired relation
$|M_1|= 5/3 \tan^2\theta_W M_2 $ and the masses of the down-type sfermions
by the SU(2) relation. All masses are given in GeV. 
}
\label{ScenarioA}
\end{table}

%%%%%%%%%%%%%%%%%%%%%%%%%%%%%%%%%%%%%%%%%%%%%%%%%%%%%%%%%%%%%%%%%%%%%%%
\subsection{\boldmath CP asymmetry for
 $\tilde{\chi}^+_1\tilde{\chi}^-_1$ production and $\tilde{\chi}^+_1$ decay}
%%%%%%%%%%%%%%%%%%%%%%%%%%%%%%%%%%%%%%%%%%%%%%%%%%%%%%%%%%%%%%%%%%%%%%%

\begin{figure}[t]
\centering
\epsfig{file=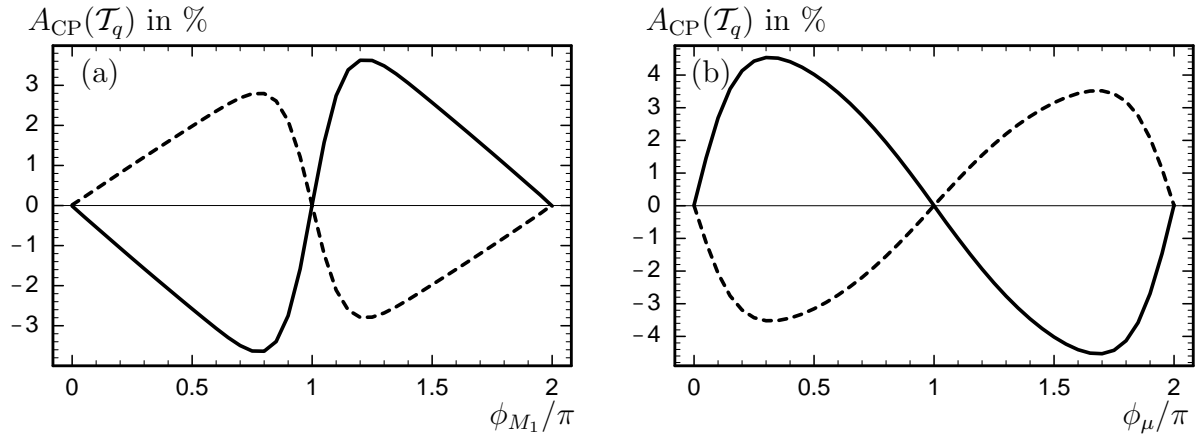,scale=0.99}
\caption{\label{fig:At11}
CP asymmetry $A_{\rm CP}(\mathcal{T}_q)$, Eq.~(\ref{ACP}),
for $e^+e^- \to \tilde{\chi}^+_1\tilde{\chi}^-_1$
with subsequent decay
$\tilde{\chi}^+_1 \to \tilde{\chi}^0_1 \bar{s} c$
for the parameters defined in Table~\ref{ScenarioA}
(a) with $\phi_{\mu}=0$ and
(b) with $\phi_{M_1}=\pi$, for $\sqrt{s}=500$~GeV and
for the beam polarizations $(P_{e^-},P_{e^+}) = (-0.8, +0.6)$ (solid),
$(P_{e^-},P_{e^+}) = (+0.8, -0.6)$ (dashed).}
\end{figure}

In the case of pair production, $e^+ e^- \to \tilde{\chi}^+_1\tilde{\chi}^-_1$,
only CP-violating couplings from the decay 
(second term in Eq.~(\ref{CPinT})) can give
rise to a CP-violating effect, because in the
production (first term in Eq.~(\ref{CPinT})) only the
absolute squares of the couplings enter.
Thus, the CP asymmetry $A_{\rm CP}(\mathcal{T}_q)$
is sensitive to the CP violation in the decay, due to the phases of 
$\mu$ and $M_1$.
Fig.~\ref{fig:At11} (a) shows the asymmetry $A_{\rm CP}(\mathcal{T}_q)$
as a function of the phase $\phi_{M_1}$ for scenario A
(see Table \ref{ScenarioA}) for $\phi_\mu=0$. 
The masses of the squarks are chosen to be
$m_{\tilde{c}}=500$~GeV and $m_{\tilde{s}}=505.9$~GeV.
The centre-of-mass energy $\sqrt{s}=500$~GeV and the two sets of 
longitudinal $e^\pm$ beam polarizations
are fixed in our study at $(P_{e^-},P_{e^+})=(+0.8,-0.6)$ and 
$(P_{e^-},P_{e^+})=(-0.8,+0.6)$.
The CP asymmetry reaches its largest value of
about $3.7\%$ for $(P_{e^-},P_{e^+})=(-0.8,+0.6)$ at
$\phi_{M_1}=1.2\pi$.
Note that the asymmetry changes its sign for
the two different sets of beam polarization due to the 
prefactor (Eq.~(\ref{prefactor}), Appendix A) which depends on the longitudinal beam
polarization. 
Note further that the asymmetry does not
have its largest absolute value for $\phi_{M_1}=0.5\pi,1.5\pi$.
This behaviour is due to a complex interplay of the $\phi_{M_1}$ dependence 
of the numerator and denominator of the asymmetry in Eq.~(\ref{Asyparts}).
In Fig.~\ref{fig:At11} (b) we show the dependence of the
CP asymmetry $A_{\rm CP}(\mathcal{T}_q)$ on $\phi_\mu$ for the same
scenario taking $\phi_{M_1}=\pi$.
The maximum value
of about $4.6\%$ of $A_{\rm CP}(\mathcal{T}_q)$ is reached at 
$\phi_\mu=0.3\pi$ for $(P_{e^-},P_{e^+})=(-0.8,+0.6)$.

\begin{figure}[p]
\centering
\epsfig{file=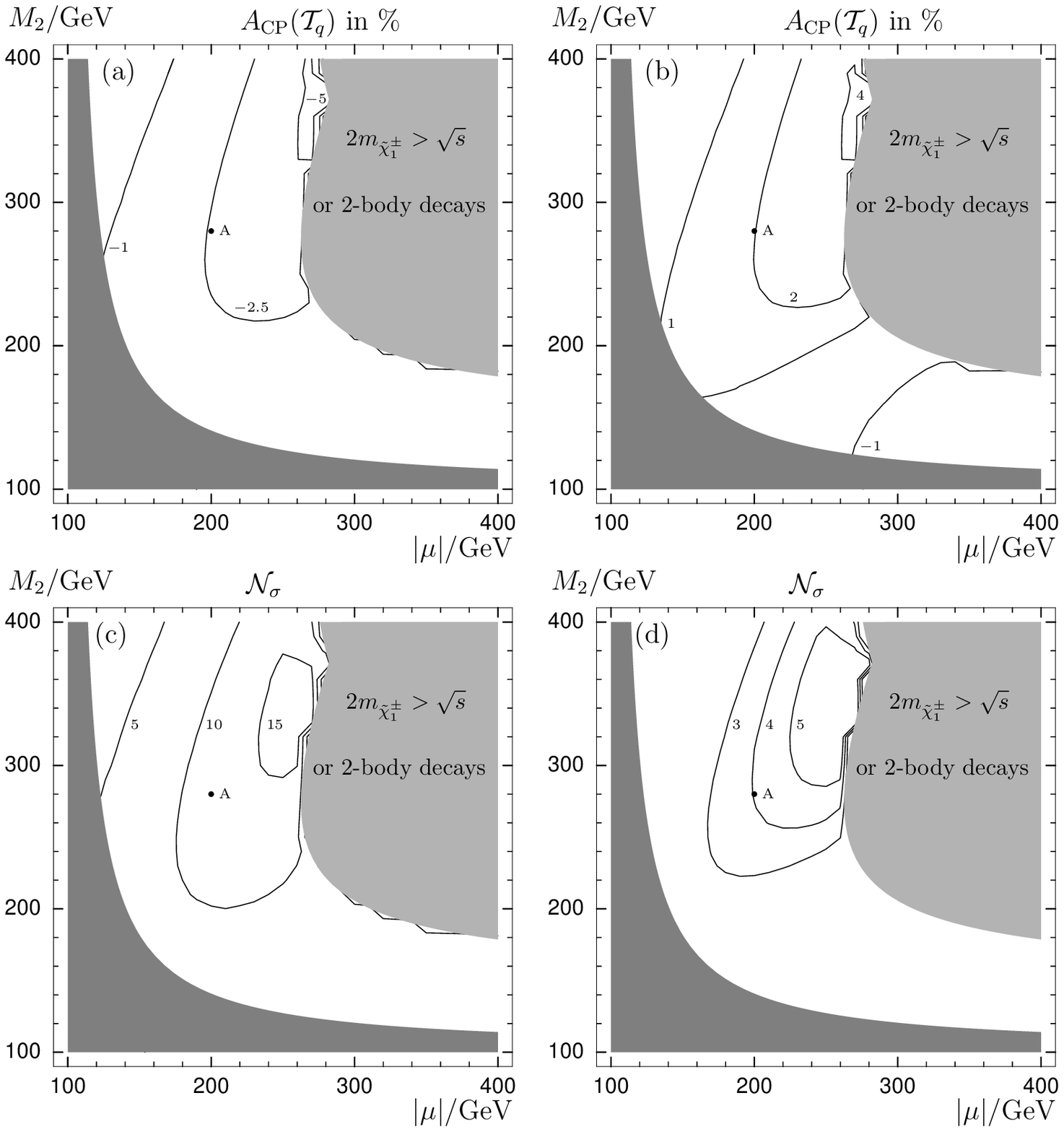,scale=0.99}
\caption{\label{fig:At11M2mu}
(a), (b) Contours of the CP asymmetry
$A_{\rm CP}(\mathcal{T}_q)$, Eq.~(\ref{ACP}), in \%
for $e^+e^- \to \tilde{\chi}^+_1\tilde{\chi}^-_1$ at $\sqrt{s}=500$~GeV
with subsequent decay
$\tilde{\chi}^+_1 \to \tilde{\chi}^0_1 \bar{s} c$
and (c), (d) contours of the number of standard deviations 
$\mathcal{N}_\sigma$, Eq.~(\ref{nsigma}), for an integrated luminosity 
$\mathcal{L}_\mathrm{int}=500~\mathrm{fb}^{-1}$, respectively.
The parameters are $\tan\beta = 5$, $m_{\tilde{\nu}} = 250$~GeV,
$m_{\tilde{c}} = 500$~GeV, $m_{\tilde{s}} = 505.9$~GeV,
$|M_1|/M_2= 5/3 \tan^2\theta_W$,
$\phi_{M_1}=0.5\pi$, $\phi_{\mu}=0$. The beam polarizations are
in (a), (c),  $(P_{e^-},P_{e^+}) = (-0.8, +0.6)$ and
in (b), (d), $(P_{e^-},P_{e^+}) = (+0.8, -0.6)$.
The point marks the scenario A, defined in Table~\ref{ScenarioA}.
In the dark-shaded area is
$m_{\tilde{\chi}^\pm_1} < 103.5$~GeV, excluded by LEP \cite{lepsusy}.
The light-shaded area shows the region that either is kinematically not accessible or
in which the three-body decay is strongly
suppressed because $m_{\tilde{\chi}^+_1} > m_W + m_{\tilde{\chi}^0_1}$.
}
\end{figure}

In Fig.~\ref{fig:At11M2mu} (a) and (b) the contours of the CP asymmetry
$A_{\rm CP}(\mathcal{T}_q)$,
Eq.~(\ref{ACP}), are shown in the $M_2$-$|\mu|$ plane.
The other MSSM parameters are chosen to be 
$\tan\beta = 5$, $m_{\tilde{\nu}} = 250$~GeV,
$m_{\tilde{c}} = 500$~GeV, $m_{\tilde{s}} = 505.9$~GeV,
$|M_1| = 5/3 \tan^2\theta_W M_2$,
$\phi_{M_1}=0.5\pi$ and $\phi_{\mu}=0$.
For both polarization configurations, $(P_{e^-},P_{e^+})=(-0.8,+0.6)$ and
$(P_{e^-},P_{e^+})=(+0.8,-0.6)$,  the absolute
value of $A_{\rm CP}(\mathcal{T}_q)$ is largest in the region
$|\mu|\approx 260$~GeV and $M_2\approx 360$~GeV with
asymmetries of about $-5\%$ $(4\%)$ for $(P_{e^-},P_{e^+})=(-0.8,+0.6)$
$((+0.8,-0.6))$. The main contributions to the numerator of 
the asymmetry are due to the
$W^+$-$\tilde{s}_L$ and $W^+$-$\tilde{c}_L$ interference 
terms.

Figs.~\ref{fig:At11M2mu} (c) and (d) show
the contours of the corresponding number of 
standard deviations $\mathcal{N}_\sigma$ for
an integrated luminosity $\mathcal{L}_\mathrm{int} = 500~\mathrm{fb}^{-1}$ 
in the $M_2$-$|\mu|$ plane.
Quite generally, the choice $(P_{e^-},P_{e^+})=(-0.8,+0.6)$ 
for longitudinal beam polarizations yields better results 
than $(P_{e^-},P_{e^+})=(+0.8,-0.6)$, because it enhances 
the sneutrino-exchange contribution to the production cross section. 
It is interesting to note
that the asymmetry $A_{\rm CP}(\mathcal{T}_q)$ is measurable with
a 5$\sigma$ significance in a large region
of the parameter space.

Our numerical results for the number of standard deviations
$\mathcal{N}_\sigma$ shown in Figs~\ref{fig:At11M2mu} (c), (d)
and Figs.~\ref{fig:At12M2mu} (c), (d) below, do not include the
influence of c-tagging, which is necessary for a measurement
of $A_{\rm CP}(\mathcal{T}_q)$. Now we want to estimate how the
detection rates are expected to be modified if the effects of 
c-tagging are also taken into account. 
Identifying the c-quark can be accomplished with the
help of vertex detectors \cite{TDR}.
It has been shown in \cite{XellaHansen:2003sw} that c-quarks will be identified
with an efficiency of about $50\%$ at a purity of $80\%$ 
in $Z^0$ decays in $e^+e^- \to q\bar{q}$ at $\sqrt{s}=500$~GeV.
Accordingly, the number of standard deviations shown in 
Figs.~\ref{fig:At11M2mu} (c), (d) and Figs.~\ref{fig:At12M2mu} (c), (d) below,
for the measurement of the CP asymmetry $A_{\rm CP}(\mathcal{T}_q)$ 
is expected to be reduced by a factor of about $0.57$.
We note that the purity of c-jets in chargino and $W$ decays is 
presumably larger \cite{Desch:2006xp}, since 
in this case fewer non-charm jets appear
(the ratio of true charm to non-charm jets is approximately 1/3
for W decays as compared to approximately 1/5 for Z decays \cite{Yao:2006px}).
For measuring the CP asymmetry $A_{\rm CP}(\mathcal{T}_q)$ it
is also necessary to distinguish the c-quark in the
decay $\tilde{\chi}^+_1 \to \tilde{\chi}^0_1 c \bar{s}$ from
its antiquark $\bar{c}$. This can be achieved with very
good precision in the semi-leptonic decays of the charmed 
hadrons. For the majority of c-jets it can also be accomplished
by the reconstruction of the vertex charge in the cases
where the charmed hadrons decay non-leptonically 
\cite{TDR,Damerell,Abe:2001dr}. The electric charge of the
c-quark can also be indirectly identified in the cases where
the second chargino, $\tilde{\chi}^-_1$, decays leptonically or
where the sign of the charge of the $\bar{c}$-jet in the decay
$\tilde{\chi}^-_1 \to \tilde{\chi}^0_1 \bar{c} s$
is determined.
   
\begin{table}[t]
\begin{center}
\begin{tabular}{|c|c|} \hline
\multicolumn{2}{|c|}{scenario B}\\
\hline\hline
$M_2$             & 150 \\ \hline
$|\mu|$           & 320 \\ \hline
$\tan\beta$       & 5   \\ \hline
$m_{\tilde{\nu}}$ & 250 \\ \hline
$m_{\tilde{u}_L}$ & 500 \\ \hline
\end{tabular}
\begin{tabular}{|c|c||c|c|c|c|c|c|} \hline
$\phi_{\mu}$ & $\phi_{M_1}$ &
$m_{\tilde{\chi}^0_1}$ & $m_{\tilde{\chi}^0_2}$ & $m_{\tilde{\chi}^0_3}$ 
& $m_{\tilde{\chi}^0_4}$  
&
$m_{\tilde{\chi}^\pm_1}$ & $m_{\tilde{\chi}^\pm_2}$\\ \hline\hline
0 &            0    & 70.6 & 132.5 & 325.7 & 347.8 & 131.0 & 347.4 \\
0 & $\frac{\pi}{2}$ & 73.4 & 132.0 & 326.2 & 347.0 & 131.0 & 347.4 \\
0 & $ \pi   $       & 76.1 & 131.4 & 326.6 & 346.2 & 131.0 & 347.4 \\ \hline
$\frac{\pi}{2}$ &           0     & 73.7 & 140.1 & 327.3 & 342.7 & 139.7 &
344.0  \\
$\frac{\pi}{2}$ & $\frac{\pi}{2}$ & 76.1 & 139.8  & 328.0 & 341.6 & 139.7 &
344.0  \\
$\frac{\pi}{2}$ & $\pi$           & 73.5 & 140.0 & 328.1 & 341.9 & 139.7  &
344.0  \\\hline
$\pi$ &       0         & 76.1 & 148.0 & 332.5 & 333.6 & 148.3  & 340.3 \\
$\pi$ & $\frac{\pi}{2}$ & 73.8 & 148.0 & 332.7 & 334.0 & 148.3  & 340.3 \\
$\pi$ & $\pi$           & 71.5 & 148.0 & 332.8 & 334.4 & 148.3  & 340.3  \\
\hline
\end{tabular}
\end{center}
\caption{Input parameters $M_2$, $|\mu|$, $\tan\beta$,
$m_{\tilde{\nu}}$ and $m_{\tilde{u}_L}=m_{\tilde{c}_L}$
and neutralino and chargino masses for scenario B
for different values of $\phi_\mu$ and 
$\phi_{M_1}$. 
$|M_1|$ is fixed by the GUT-inspired relation
$|M_1|= 5/3 \tan^2\theta_W M_2 $ and the masses of the down-type sfermions
by the SU(2) relation. All masses are given in GeV. 
}
\label{ScenarioB}
\end{table}

%%%%%%%%%%%%%%%%%%%%%%%%%%%%%%%%%%%%%%%%%%%%%%%%%%%%%%%%%%%%%%%%%%%%%
\subsection{\boldmath CP-odd asymmetry for
 $\tilde{\chi}^+_1\tilde{\chi}^-_2$ production and $\tilde{\chi}^+_1$ decay}
%%%%%%%%%%%%%%%%%%%%%%%%%%%%%%%%%%%%%%%%%%%%%%%%%%%%%%%%%%%%%%%%%%%%%
Now we consider the production process 
$e^+e^- \to \tilde{\chi}^+_1\tilde{\chi}^-_2$ at $\sqrt{s}=500$~GeV with
subsequent decays of the $\tilde{\chi}^+_1$. 
In this case $A_{\rm CP}(\mathcal{T}_q)$ is sensitive to the
CP-violating couplings in the production and decay amplitudes
(i.e.\ it is sensitive to both terms in (\ref{CPinT})). 

%%%%%%%%%%%%%%%%%%%%%%%%%%%%%%%%%%%%%%%%%%%%%%%%%%%%%
\subsubsection{\boldmath Hadronic decay
 $\tilde{\chi}^+_1 \to \tilde{\chi}^0_1 \bar{s} c$}
%%%%%%%%%%%%%%%%%%%%%%%%%%%%%%%%%%%%%%%%%%%%%%%%%%%%%
In the case of hadronic decays, $\tilde{\chi}^+_1 \to \tilde{\chi}^0_1 \bar{s} c$, $c$-charge 
tagging
is highly desirable because of the complicated cascade decays of the heavy chargino. 

\begin{figure}[t]
\centering
\epsfig{file=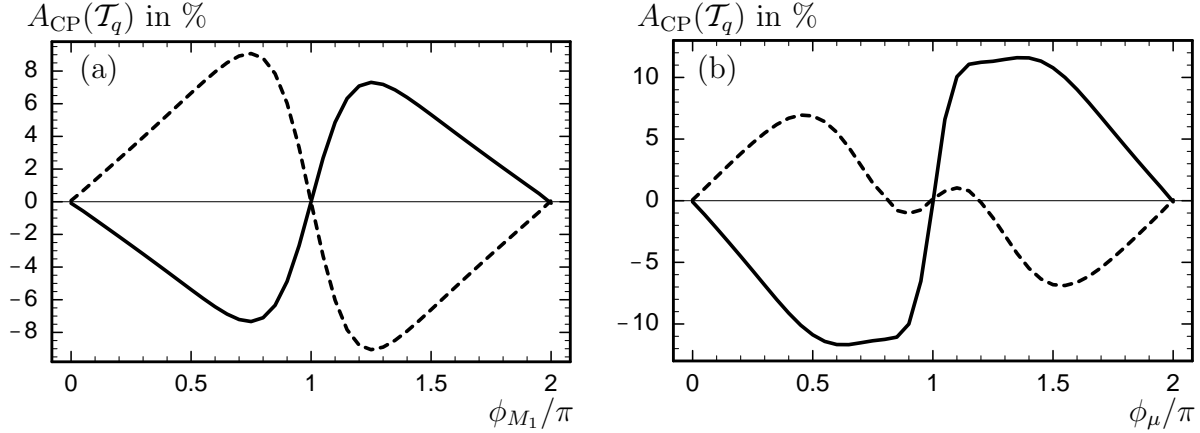,scale=0.99}
\caption{\label{fig:At12}
CP asymmetry $A_{\rm CP}(\mathcal{T}_q)$, Eq.~(\ref{ACP}), 
for $e^+e^- \to \tilde{\chi}^+_1\tilde{\chi}^-_2$
with subsequent decay
$\tilde{\chi}^+_1 \to \tilde{\chi}^0_1 \bar{s} c$
for the parameters given in Table~\ref{ScenarioB} 
(a) with $\phi_{\mu}=0$ and (b) with $\phi_{M_1}=0$, 
for $\sqrt{s}=500$~GeV and for the beam polarizations
$(P_{e^-},P_{e^+}) = (-0.8, +0.6)$ (solid),
$(P_{e^-},P_{e^+}) = (+0.8, -0.6)$ (dashed).}
\end{figure}

In Fig.~\ref{fig:At12} (a) we show the CP asymmetry
$A_{\rm CP}(\mathcal{T}_q)$, Eq.~(\ref{ACP}), as a function
of $\phi_{M_1}$ for scenario B given in Table~\ref{ScenarioB},
with $\phi_\mu=0$. The longitudinal beam polarization
is $(P_{e^-},P_{e^+})=(-0.8,+0.6)$
$((+0.8,-0.6))$. The asymmetry 
reaches its largest value of about $9\%$ $(7\%)$ for
$\phi_{M_1}=0.7\pi~(1.2\pi)$.
Fig.~\ref{fig:At12} (b) shows $A_{\rm CP}(\mathcal{T}_q)$
as a function of $\phi_\mu$ for $\phi_{M_1}=0$.
The largest value of the CP asymmetry is reached at $\phi_\mu=1.4\pi$ $(0.6\pi)$.
Note that the asymmetry can be large ($\sim 10$\%), even
for values of $\phi_\mu$ close to $\pi$.
As can be seen in Figs.~\ref{fig:At12} (a) and (b), it
changes the sign for the two choices of beam polarizations.

\begin{figure}[p]
\centering
\epsfig{file=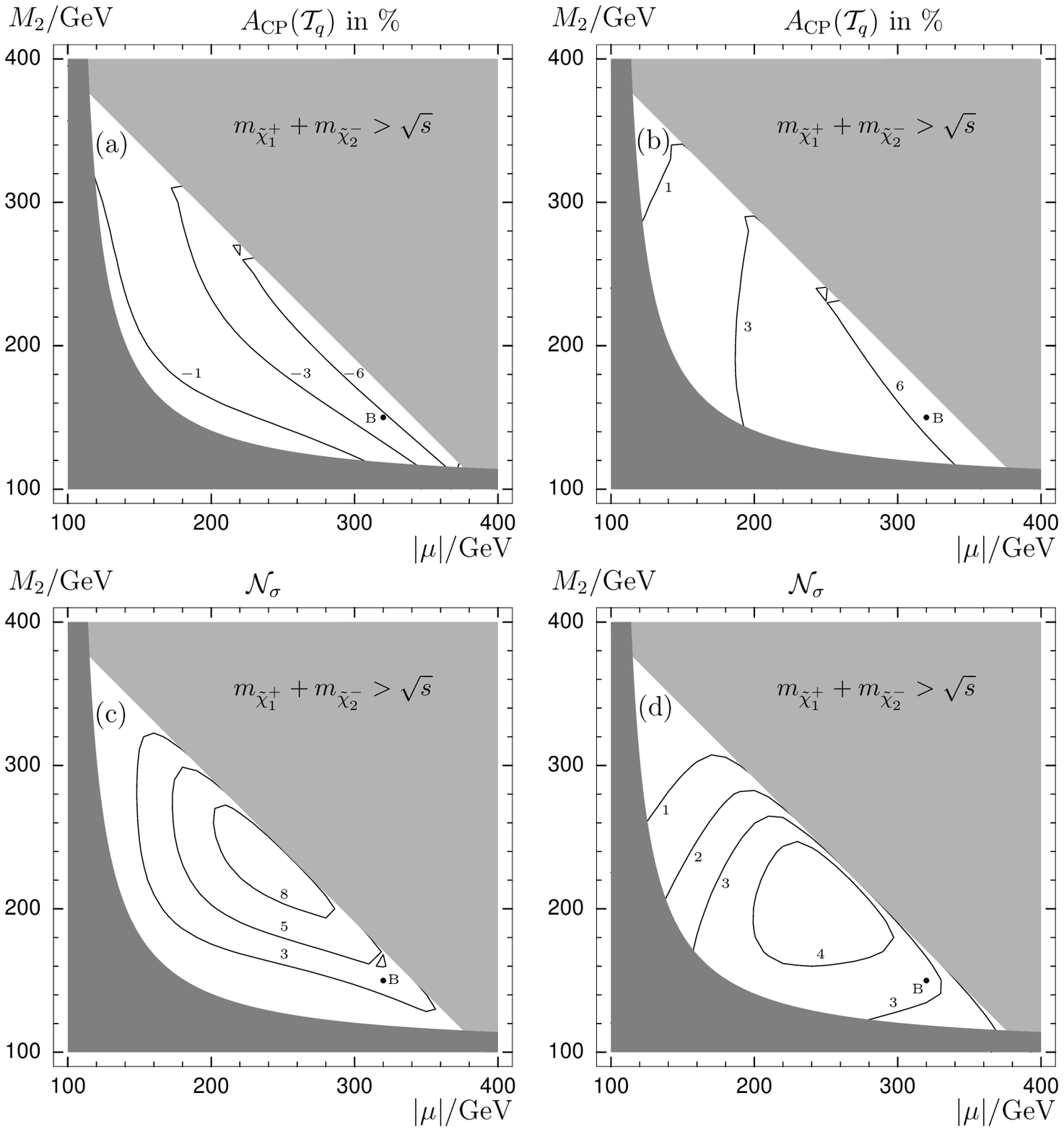,scale=0.99}
\caption{\label{fig:At12M2mu}
(a), (b) Contours of the CP asymmetry
$A_{\rm CP}(\mathcal{T}_q)$, Eq.~(\ref{ACP}),
in \% for $e^+e^- \to \tilde{\chi}^+_1\tilde{\chi}^-_2$ at $\sqrt{s}=500$~GeV
with subsequent decay
$\tilde{\chi}^+_1 \to \tilde{\chi}^0_1 \bar{s} c$
and (c), (d) contours of the number of standard deviations 
$\mathcal{N}_\sigma$, Eq.~(\ref{nsigma}), for an integrated luminosity 
$\mathcal{L}_\mathrm{int}=500~\mathrm{fb}^{-1}$, respectively.
The parameters are $\tan\beta = 5$, $m_{\tilde{\nu}} = 250$~GeV,
$m_{\tilde{c}} = 500$~GeV, $m_{\tilde{s}} = 505.9$~GeV,
$|M_1|/M_2= 5/3 \tan^2\theta_W$,
$\phi_{M_1}=0.5\pi$ and $\phi_{\mu}=0$.
The beam polarizations are in 
(a), (c) $(P_{e^-},P_{e^+}) = (-0.8, +0.6)$ and in
(b), (d) $(P_{e^-},P_{e^+}) = (+0.8, -0.6)$.
The point marks the scenario B, 
defined in Table~\ref{ScenarioB}.
In the dark-shaded area is
$m_{\tilde{\chi}^\pm_1} < 103.5$~GeV, excluded by LEP \cite{lepsusy}.
The light-shaded area is kinematically not accessible.
}
\end{figure}

In Fig.~\ref{fig:At12M2mu} (a) and (b)
the contours of 
$A_{\rm CP}(\mathcal{T}_q)$, Eq.~(\ref{ACP}),
are shown in the $M_2$-$|\mu|$
plane for $\tan\beta = 5$, $m_{\tilde{\nu}} = 250$~GeV,
$m_{\tilde{c}} = 500$~GeV, $m_{\tilde{s}} = 505.9$~GeV,
$|M_1|= 5/3 \tan^2\theta_W M_2$,
$\phi_{M_1}=0.5\pi$ and $\phi_{\mu}=0$.
Figs.~\ref{fig:At12M2mu} (c) and (d) show 
the corresponding contours for $\mathcal{N}_\sigma$, Eq.~(\ref{nsigma}),
for $\mathcal{L}_\mathrm{int} = 500~\mathrm{fb}^{-1}$
in the $M_2$-$|\mu|$ plane. Also in this case the choice
$(P_{e^-},P_{e^+})=(-0.8,+0.6)$ 
enhances the statistical significance for a measurement of
$A_\mathrm{CP}(\mathcal{T}_q)$.

%%%%%%%%%%%%%%%%%%%%%%%%%%%%%%%%%%%%%%%%%%%%%%%%%%%%%%%
\subsubsection{\boldmath Leptonic decay
 $\tilde{\chi}^+_1 \to \tilde{\chi}^0_1 \ell^+ \nu$}
%%%%%%%%%%%%%%%%%%%%%%%%%%%%%%%%%%%%%%%%%%%%%%%%%%%%%%%

\begin{table}[t]
\begin{center}
\begin{tabular}{|c|c|} \hline
\multicolumn{2}{|c|}{scenario C}\\
\hline\hline
$M_2$             & 120 \\ \hline
$|\mu|$           & 320 \\ \hline
$\tan\beta$       & 5   \\ \hline
$m_{\tilde{\nu}}$ & 250 \\ \hline
$m_{\tilde{u}_L}$ & 500 \\ \hline
\end{tabular}
\begin{tabular}{|c|c||c|c|c|c|c|c|} \hline
$\phi_{\mu}$ & $\phi_{M_1}$ &
$m_{\tilde{\chi}^0_1}$ & $m_{\tilde{\chi}^0_2}$ & $m_{\tilde{\chi}^0_3}$ 
& $m_{\tilde{\chi}^0_4}$  
&
$m_{\tilde{\chi}^\pm_1}$ & $m_{\tilde{\chi}^\pm_2}$\\ \hline\hline
0 &            0    & 55.9 & 105.5 & 326.1 & 344.8 & 104.2 & 344.8 \\
0 & $\frac{\pi}{2}$ & 58.6 & 105.1 & 326.4 & 344.1 & 104.2 & 344.8 \\
0 & $ \pi   $       & 61.4 & 104.5 & 326.8 & 343.5 & 104.2 & 344.8 \\ \hline
$\frac{\pi}{2}$ &           0     & 59.1 & 112.7 & 327.5 & 340.6 & 112.5 &
342.2  \\
$\frac{\pi}{2}$ & $\frac{\pi}{2}$ & 61.3 & 112.5  & 328.0 & 339.7 & 112.5 &
342.2  \\
$\frac{\pi}{2}$ & $\pi$           & 58.7 & 112.8 & 328.1 & 340.1 & 112.5  &
342.2  \\\hline
$\pi$ &       0         & 61.2 & 120.2 & 331.9 & 333.3 & 120.5  & 339.4 \\
$\pi$ & $\frac{\pi}{2}$ & 59.1 & 120.2 & 331.5 & 334.1 & 120.5  & 339.4 \\
$\pi$ & $\pi$           & 56.8 & 120.2 & 331.2 & 334.7 & 120.5  & 339.4  \\
\hline
\end{tabular}
\end{center}
\caption{Input parameters $M_2$, $|\mu|$, $\tan\beta$,
$m_{\tilde{\nu}}$ and $m_{\tilde{u}_L}=m_{\tilde{c}_L}$ and
neutralino and chargino masses for scenario C
for different values of $\phi_\mu$ and 
$\phi_{M_1}$. 
$|M_1|$ is fixed by the GUT-inspired relation
$|M_1|= 5/3 \tan^2\theta_W M_2 $ and the masses of the down-type sfermions
by the SU(2) relation. All masses are given in GeV. 
}
\label{ScenarioC}
\end{table}

\begin{figure}[t]
\centering
\epsfig{file=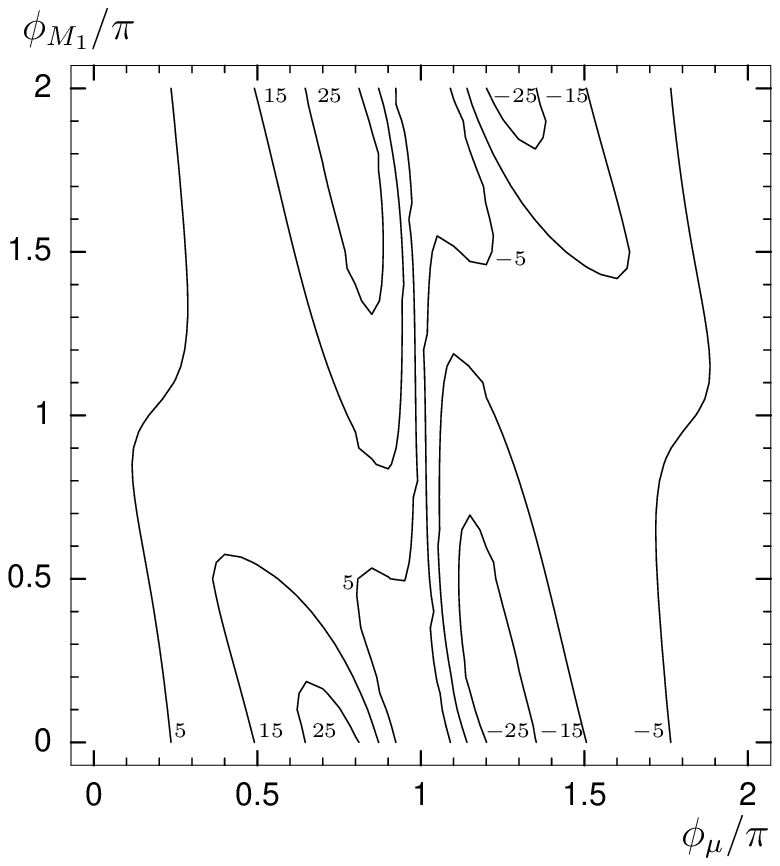}
\caption{\label{fig:AtL12phiM1phimu}
Contours of the CP asymmetry $A_{\rm CP}(\mathcal{T}_\ell)$, Eq.~(\ref{ACP}), in \%
for $e^+e^- \to \tilde{\chi}^+_1\tilde{\chi}^-_2$ at $\sqrt{s}=500$~GeV
with subsequent decay
$\tilde{\chi}^+_1 \to \tilde{\chi}^0_1 \ell^+ \nu$
for the parameters defined in Table~\ref{ScenarioC} and
$(P_{e^-},P_{e^+}) = (-0.8, +0.6)$.}
\end{figure}

In this section we analyse the CP asymmetry 
$A_{\rm CP}(\mathcal{T}_\ell)$, Eq.~(\ref{ACP}), based on the triple product correlation 
$\mathcal{T}_\ell=\vec{p}_{\ell^+}
\cdot (\vec{p}_{e^-} \times \vec{p}_{\tilde{\chi}^+_1})$. 
For the process $e^+ e^- \to \tilde{\chi}^+_1\tilde{\chi}^-_2$ 
the asymmetry $A_{\rm CP}(\mathcal{T}_\ell)$ is only sensitive to 
CP-violating couplings in the production amplitude, which are involved 
in the first term of (\ref{CPinT}),
because the CP sensitive couplings in the decay
(c.f.\ Eqs.~(\ref{eq_te1})--(\ref{eq_te4})) do not contain the triple product 
$\mathcal{T}_\ell$.
This means $A_{\rm CP}(\mathcal{T}_\ell)$
is proportional to $\sin(\phi_\mu)$ and therefore 
$A_{\rm CP}(\mathcal{T}_\ell) \equiv 0$ for $\phi_\mu = 0, \pi, 2\pi, \dots$,
independently of $\phi_{M_1}$.
Hence, by measuring the CP asymmetries $A_{\rm CP}(\mathcal{T}_\ell)$ and
$A_{\rm CP}(\mathcal{T}_q)$ one can separately study the influence of $\phi_\mu$
and $\phi_{M_1}$.

In Fig.~\ref{fig:AtL12phiM1phimu} we show the contour lines of
the CP-odd asymmetry $A_{\rm CP}(\mathcal{T}_\ell)$, Eq.~(\ref{ACP}),
for scenario C of Table~\ref{ScenarioC} in the $\phi_{M_1}$-$\phi_{\mu}$
plane. 
Fig.~\ref{fig:AtL12phiM1phimu} illustrates that the asymmetry 
$A_{\rm CP}(\mathcal{T}_\ell)$ can be large for values of $\phi_\mu$ close to $\pi$.
For instance, for $\phi_{M_1}=1.5\pi$ and $\phi_\mu=0.9\pi$
one obtains an asymmetry of about $23\%$. However, the corresponding
cross section is only about $0.16$~fb. 

\begin{figure}[t]
\centering
\epsfig{file=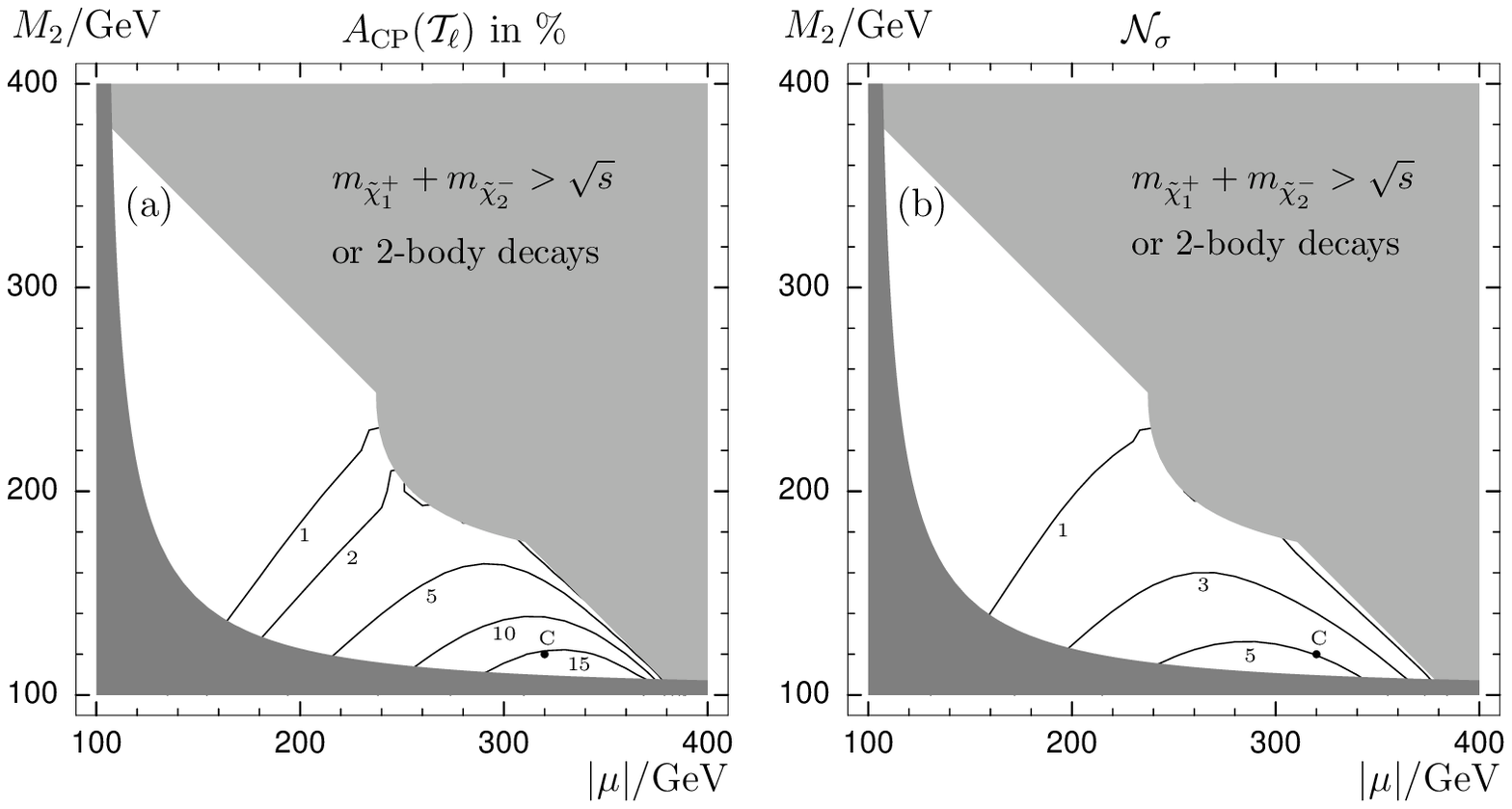,scale=0.99}
\caption{\label{fig:AtL12M2mu}
(a) Contours of the CP asymmetry 
$A_{\rm CP}(\mathcal{T}_\ell)$, Eq.~(\ref{ACP}), in \%
for $e^+e^- \to \tilde{\chi}^+_1\tilde{\chi}^-_2$ at $\sqrt{s}=500$~GeV
with subsequent decay
$\tilde{\chi}^+_1 \to \tilde{\chi}^0_1 \ell^+ \nu$
and (b) contours of the number of standard deviations
$\mathcal{N}_\sigma$, Eq.~(\ref{nsigma}), for an integrated luminosity 
$\mathcal{L}_\mathrm{int}=500~\mathrm{fb}^{-1}$, respectively.
The parameters are $\tan\beta = 5$, $m_{\tilde{\nu}} = 250$~GeV,
$m_{\tilde{\ell}} = 261.7$~GeV,
$|M_1|/M_2= 5/3 \tan^2\theta_W$,
$\phi_{M_1}=0$, $\phi_{\mu}=0.5\pi$
and the beam polarizations are $(P_{e^-},P_{e^+}) = (-0.8, +0.6)$.
The point marks the scenario C, 
defined in Table~\ref{ScenarioC}.
In the dark-shaded area is
$m_{\tilde{\chi}^\pm_1} < 103.5$~GeV, excluded by LEP \cite{lepsusy}.
The light-shaded area shows the region that either is not kinematically accessible or
in which the three-body decay is strongly
suppressed because $m_{\tilde{\chi}^+_1} > m_W + m_{\tilde{\chi}^0_1}$.
}
\end{figure}

In Fig.~\ref{fig:AtL12M2mu} (a) and (b),
the CP asymmetry
$A_{\rm CP}(\mathcal{T}_\ell)$, Eq.~(\ref{ACP}), and the number
of standard deviations $\mathcal{N}_\sigma$, Eq.~(\ref{nsigma}),
are shown for $\mathcal{L}_\mathrm{int}=500~\mathrm{fb}^{-1}$, respectively,
in the \mbox{$M_2$-$|\mu|$} plane. The MSSM parameters are 
$\tan\beta = 5$, $m_{\tilde{\nu}} = 250$~GeV,
$m_{\tilde{\ell}} = 261.7$~GeV,
$|M_1| = 5/3 \tan^2\theta_W M_2$,
$\phi_{M_1}=0$ and $\phi_{\mu}=0.5\pi$.
The asymmetry reaches its largest values of about $15\%$ 
in gaugino-like scenarios.
For example, for scenario C, $A_{\rm CP}(\mathcal{T}_\ell)$ 
can be measured with a $5 \sigma$ significance.

In order to be able to measure $A_{CP}(\mathcal{T}_{\ell})$,
the production plane has to be reconstucted. Provided the masses
of the particles are known, this could be accomplished depending
on the decay pattern of $\tilde{\chi}^-_2$ \cite{Bartl:2005uh}. 
For example, in the case
of scenario C (fixing $\phi_\mu=\pi$ and $\phi_{M_1}=0$) the decays
of $\tilde{\chi}^-_2$ which can be used for the reconstruction of the
production plane are (i) $\tilde{\chi}^-_2 \to \tilde{\chi}^0_2 W^-$,
$\tilde{\chi}^0_2 \to \tilde{\chi}^0_1q\bar{q}(\tilde{\chi}^0_1\ell \bar{\ell})$,
$W^- \to q\bar{q}'$, (ii) $\tilde{\chi}^-_2 \to \tilde{\chi}^-_1 Z^0$,
$\tilde{\chi}^-_1 \to \tilde{\chi}^0_1q\bar{q}'$,
$Z^0 \to q\bar{q}(\ell\bar{\ell})$ and (iii) 
$\tilde{\chi}^-_2 \to \tilde{\chi}^-_1 h^0$,
$\tilde{\chi}^-_1 \to \tilde{\chi}^0_1 q\bar{q}'$,
$h^0 \to q\bar{q}(\ell\bar{\ell})$.
The masses of the particles involved are given in Table \ref{ScenarioC},
and we take $m_{h^0}=115$~GeV. In the decay chains (i) ((ii), (iii)) we
obtain two invariant mass constraints from the on-shell $\tilde{\chi}^0_2$
($\tilde{\chi}^-_1$) and the only invisible particle in the final
states of the decay chains (i)-(iii) is $\tilde{\chi}^0_1$.
In these cases the production plane can be reconstructed
up to a twofold ambiguity.
The branching ratios of the decay chains (i), (ii) and (iii) are $20\%$,
$25\%$ and $17\%$, respectively. This means that in this case 
about $62\%$ of the
decays of $\tilde{\chi}^-_2$ can be used for the reconstruction of the
production plane, which implies that the number
of standard deviations $\mathcal{N}_\sigma$ shown
in Fig.~\ref{fig:AtL12M2mu} (b) would have to be reduced accordingly.
We note that in scenario C the decays 
$\tilde{\chi}^-_2 \to \bar{\tilde{\nu}}\ell^-$ and
$\tilde{\chi}^-_2 \to \tilde{\ell}^-_L \bar{\nu}$ are suppressed
and the decay $\tilde{\chi}^-_2 \to \tilde{s}\bar{c}$
is kinematically not accessible.
In the case that these decays contribute significantly, then it is 
again possible to reconstruct the
production plane in the decays $\tilde{\chi}^-_2 \to \bar{\tilde{\nu}}\ell^-$
and $\tilde{\chi}^-_2 \to \tilde{s}\bar{c}$.

In order to predict the significance more accurately, 
detailed Monte Carlo analysis including detector simulations
and particle identification and reconstruction efficiencies would
be required, which is, however,
beyond the scope of the present work. For instance, a Monte Carlo
analysis for a CP asymmetry in the production and decay of neutralinos
with longitudinal beam polarization has been carried out in 
\cite{Aguilar-Saavedra:2004dz}.

%%%%%%%%%%%%%%%%%%%%%%%%%%%%%%%%%%%%%%%%%%%
\section{Summary and conclusions}
%%%%%%%%%%%%%%%%%%%%%%%%%%%%%%%%%%%%%%%%%%%

We have analysed CP-sensitive observables
in chargino production $e^+e^-\to \tilde{\chi}^+_1\tilde{\chi}^-_{1,2}$
with subsequent hadronic and leptonic three-body decays
$\tilde{\chi}^+_1 \to \tilde{\chi}^0_1 \bar{f}^d f^u$
($f^d=e,\mu,s\,$ and $f^u=\nu_e,\nu_\mu,c$)
at an $e^+ e^-$ linear collider with centre-of-mass
energy $\sqrt{s}=500$~GeV, integrated luminosity
$\mathcal{L}_{\rm int}=500$ fb$^{-1}$ and longitudinally polarized beams.
Our framework has been the MSSM with complex parameters.
We have constructed CP-odd asymmetries 
with the help of triple product correlations 
between the momenta of the incoming 
and outgoing particles. 

Considering the production process $e^+e^-\to \tilde{\chi}^+_1\tilde{\chi}^-_1$
followed by the hadronic three-body decay  
$\tilde{\chi}^+_1 \to \tilde{\chi}^0_1 \bar{s} c$,
we have defined the CP asymmetry $A_{\rm CP}(\mathcal{T}_q)$ 
that is based on the triple product $\mathcal{T}_q=\vec{p}_{\bar{s}}
\cdot (\vec{p}_{c}\times\vec{p}_{e^-})$.
The asymmetry $A_{\rm CP}(\mathcal{T}_q)$ is sensitive to CP violation
in the decay and depends on the phases $\phi_\mu$ and $\phi_{M_1}$ appearing in the
chargino/neutralino system. We have shown that the measurability
of the asymmetry $A_{\rm CP}(\mathcal{T}_q)$ can be significantly increased
by a suitable choice of beam polarizations. 
Choosing $(P_{e^-},P_{e^+})=(-0.8,+0.6)$,
$A_{\rm CP}(\mathcal{T}_q)$ can be probed at the $5 \sigma$ level
in a large region of the MSSM parameter space. 

For the production process $e^+e^-\to \tilde{\chi}^+_1\tilde{\chi}^-_2$
we have separately considered the hadronic three-body decay
$\tilde{\chi}^+_1 \to \tilde{\chi}^0_1 \bar{s} c$ and
the leptonic three-body decays
$\tilde{\chi}^+_1 \to \tilde{\chi}^0_1 \ell^+ \nu$, $\ell=e,\mu$.
For the hadronic three-body decay we have studied again
the CP asymmetry that is based on the triple product $\mathcal{T}_q$.
In this case, the resulting CP asymmetry
is sensitive to CP violation in production and decay.
Also this asymmetry can be probed at the $5 \sigma$ 
level for MSSM parameters with appreciable gaugino-higgsino mixing. 
For the leptonic three-body decays, we have studied
the asymmetry  $A_{\rm CP}(\mathcal{T}_\ell)$ that is based on
the triple product $\mathcal{T}_\ell=\vec{p}_{\ell^+}
\cdot (\vec{p}_{e^-} \times \vec{p}_{\tilde{\chi}^+_i})$,which
is sensitive to CP violation in the production only and hence
to the phase $\phi_\mu$.
We have found that the measurability of $A_{\rm CP}(\mathcal{T}_\ell)$
is somewhat decreased with respect to the previously considered asymmetries; however,
in some regions of the MSSM parameter space
it is accessible at the $3 \sigma$ level.
As the two types of CP-odd asymmetries are sensitive to 
various combinations of the phases $\phi_\mu$ and 
$\phi_{M_1}$, their measurement will allow CP violation to be tested
in the chargino/neutralino sector. 
Moreover, we have demonstrated that the CP-odd asymmetries
studied in this paper can be large even for small
CP-violating phases $\phi_\mu$ and 
$\phi_{M_1}$, which are favoured by the EDM constraints. 
  
%%%%%%%%%%%%%%%%%%%%%%%%%%%%%%%%%%%%%%%%%%%
\section*{Acknowledgements}
%%%%%%%%%%%%%%%%%%%%%%%%%%%%%%%%%%%%%%%%%%%

We thank O.~Kittel, W.~Majerotto and H.-U.~Martyn 
for valuable discussions.
This work is supported by the `Fonds zur F\"orderung der
wissenschaftlichen Forschung' (FWF) of Austria, project No. P18959-N16,
and by the German Federal Ministry of Education and Research (BMBF) 
under contract number 05HT4WWA/2.
The authors acknowledge support from EU under the 
MRTN-CT-2006-035505 network programme.

%%%%%%%%%%%%%%%%%%%%%%%%%%%%%%%%%%%%%%%%
\begin{appendix}
%%%%%%%%%%%%%%%%%%%%%%%%%%%%%%%%%%%%%%%%

%%%%%%%%%%%%%%%%%%%%%%%%%%%%%%%%%%%%%%%%%%%%%%%%%%%%%%%%
\section{Formalism}
%%%%%%%%%%%%%%%%%%%%%%%%%%%%%%%%%%%%%%%%%%%%%%%%%%%%%%%%

The full expressions for the terms $\Sigma^a_P(\tilde{\chi}^+_i)$ 
and $\Sigma^a_D(\tilde{\chi}^+_i)$ are 
given in \cite{Moortgat-Pick:1998sk}. In the following we decompose
$\Sigma^a_P(\tilde{\chi}^+_i)$ 
and $\Sigma^a_D(\tilde{\chi}^+_i)$ into 
T-odd and the T-even terms, which are needed in subsection
\ref{subasy}.

%%%%%%%%%%%%%%%%%%%%%%%%%%%%%%%%%%%%%%%%%%%%%%%%%%%%%%%%%%
\subsection{T-odd terms of production and T-even terms of decay}
%%%%%%%%%%%%%%%%%%%%%%%%%%%%%%%%%%%%%%%%%%%%%%%%%%%%%%%%%%
In the definition of the momenta 
and polarization 4-vectors we follow \cite{Bartl:2004jj,Moortgat-Pick:1998sk}.
$p_i$, $i=1,\dots,7$, are the 4-momenta of
the particles $e^-$, $e^+$, $\tilde\chi^+_i$, $\tilde\chi^-_j$,
$\tilde\chi^0_k$, $f^d$ and $f^u$, respectively, see Figs.~\ref{Fig:FeynProd}
and \ref{Fig:FeynDecay}.
It can be shown 
that all contributions to the T-odd terms 
$\Sigma^{a,\mathrm{O}}_P(\tilde{\chi}^+_i)$ 
in Eq.~(\ref{CPinT}) 
contain a factor
\begin{equation}\label{eq_f5_a=2}
f_5^{a} = i \cdot m_j \epsilon_{\mu\nu\rho\sigma} p_2^{\mu}
p_1^{\nu} s^{a,\rho} p_3^{\sigma}
\end{equation}
which vanishes for longitudinal polarization ($a=3$) and transverse
polarization in the production plane ($a=1$)
so that 
we have only to include the spin terms for transverse polarization
of the chargino $\tilde{\chi}^+_i$
perpendicular to the production plane
($a=2$):
\begin{eqnarray}
\Sigma^{2,\mathrm{O}}_P(\tilde{\chi}^+_i) & = &
\Sigma^{2,\mathrm{O}}_P(\gamma Z) +
\Sigma^{2,\mathrm{O}}_P(\gamma \tilde{\nu}) +
\Sigma^{2,\mathrm{O}}_P(Z Z) +
\Sigma^{2,\mathrm{O}}_P(Z \tilde{\nu})
\end{eqnarray}
with
\begin{eqnarray}
\Sigma^{2,\mathrm{O}}_P(\gamma Z) &=&
  g^4 \tan^2\theta_W~{\rm Re}\Big\{\Delta(\gamma)\Delta^{*}(Z)\delta_{ij}
  c^{P}_{-}(\gamma Z) (O_{ij}^{'L*}-O_{ij}^{'R*}) f^{a=2}_5  \Big\}, \label{eq_spgamZ_o}\\
\Sigma^{2,\mathrm{O}}_P(\gamma \tilde{\nu}) &=& 
  - \frac{g^4}{2}~\sin^2\theta_W~ {\rm Re}\Big\{\Delta(\gamma)
  \Delta^{*}(\tilde{\nu}) \delta_{ij}
  c^{P}_{+}(\gamma\tilde{\nu})~V^*_{i1}~V_{j1} f^{a=2}_5\Big\}, \label{eq_spgamsnu_o}\\
\Sigma^{2,\mathrm{O}}_P(ZZ) &=& 
 \frac{g^4}{2\cos^4\theta_W}~ |\Delta(Z)|^2 
  \Big[  c^{P}_{-}(ZZ)
       (O_{ij}^{'R} O_{ij}^{'L*}-O_{ij}^{'L} O_{ij}^{'R*}) f^{a=2}_5 \Big],\\
\Sigma^{2,\mathrm{O}}_P(Z \tilde{\nu}) &=& 
  - \frac{g^4}{2\cos^2\theta_W}~ {\rm Re}\Big\{\Delta(Z) \Delta^{*}(\tilde{\nu})
c^{P}_{+}(Z \tilde{\nu}) ~V^*_{i1}~V_{j1} O_{ij}^{'R} f^{a=2}_5 \Big\}. \label{eq_spZsnu_o}
\end{eqnarray}
Here
\begin{equation}\label{prefactor}
c^{P}_{\pm}(\alpha\beta)
=\pm c^{L}(\alpha)  c^{L}(\beta)(1-P_{e^-})(1+P_{e^+})
+c^{R}(\alpha) c^{R}(\beta) (1+P_{e^-})(1-P_{e^+})
\end{equation}
with
\begin{eqnarray}
c^{L}(\gamma) & = & 1, \quad
c^{L}(Z)=L_{e}, \quad
c^{L}(\tilde{\nu})=1,\\
c^{R}(\gamma) & = & 1, \quad
c^{R}(Z)=R_{e}, \quad
c^{R}(\tilde{\nu})=0,
\end{eqnarray}
and $P_{e^-}$ and $P_{e^+}$ is the degree of longitudinal polarization of
the electron beam and positron beam, respectively. The propagators are 
$\Delta(\gamma)=i/(p_1+p_2)^2, \Delta(Z)=i/((p_1+p_2)^2-m^2_Z)$ and 
$\Delta(\tilde\nu)=i/((p_1-p_4)^2-m^2_{\tilde\nu})$ and the couplings
are given by
\begin{eqnarray}
L_{f} & = & T_{3f}-e_{f}\sin^2\theta_W, \quad
 R_{f}=-e_{f}\sin^2\theta_W,\\
O_{ij}^{'L} & = & -V_{i1} V_{j1}^{*}-\frac{1}{2} V_{i2} V_{j2}^{*}+
\delta_{ij} \sin^2\theta_W,\\
O_{ij}^{'R} & = & -U_{i1}^{*} U_{j1}-\frac{1}{2} U_{i2}^{*} U_{j2}+
\delta_{ij} \sin^2\theta_W,
\end{eqnarray}
where $g$ is the weak coupling constant, $e_f$ and
$T_{3f}$ are the charge (in units of $e$) and the third component of the
weak isospin of the fermion $f$, $\theta_W$ is the weak mixing angle. 
The unitary $(2\times2)$-matrices $U$ and $V$ diagonalize
the complex chargino mass matrix, see for instance  \cite{ref13}.

Note that $f^{a}_5$ is purely imaginary,
so that, for example, $\Sigma^{2,\mathrm{O}}_P(\gamma Z)$,
Eq.~(\ref{eq_spgamZ_o}), is non-vanishing only if the 
couplings $O_{ij}^{'L,R}$ are 
complex and gives a CP-sensitive contribution
to the asymmetry $A_{\rm CP}(\mathcal{T}_{q,\ell})$.
Analogous contributions come from the
other terms in $\Sigma^{2,\mathrm{O}}_P$,
Eqs.~(\ref{eq_spgamsnu_o})--(\ref{eq_spZsnu_o}).
We have to multiply $\Sigma^{2,\mathrm{O}}_P$ in
Eq.~(\ref{CPinT}) by 
\begin{eqnarray}
\Sigma^{2,\mathrm{E}}_D(\tilde{\chi}^+_i) & = &
\Sigma^{2,\mathrm{E}}_D(W^+ W^+) +
\Sigma^{2,\mathrm{E}}_D(W^+ \tilde{f}^d_L) +
\Sigma^{2,\mathrm{E}}_D(W^+ \tilde{f}^u_L) \nonumber\\
& & {}+ \Sigma^{2,\mathrm{E}}_D(\tilde{f}^d_L \tilde{f}^d_L) +
\Sigma^{2,\mathrm{E}}_D(\tilde{f}^d_L \tilde{f}^u_L) +
\Sigma^{2,\mathrm{E}}_D(\tilde{f}^u_L \tilde{f}^u_L),
\end{eqnarray}
with
\begin{eqnarray}
\Sigma^{2,\mathrm{E}}_D(W^{+} W^{+}) &=& 
 2 g^4~|\Delta(W)|^2 \Big[2 (|O_{ki}^R|^2~g^{a=2}_2-
|O_{ki}^L|^2~g^{a=2}_1) \nonumber\\
 & & {} - (O_{ki}^{L*} O_{ki}^R +O_{ki}^L O_{ki}^{R*}) (g^{a=2}_4-g^{a=2}_3) \Big],\\
\Sigma^{2,\mathrm{E}}_D(W^{+} \tilde{f}^d_L)&=&g^4\sqrt{2}~ {\rm Re}\Big\{
  \Delta(W) \Delta^{*}(\tilde{f}^d_L)~2~
  U^*_{i1} f^L_{f^dk}\Big[2 O_{ki}^R g^{a=2}_2 \nonumber\\
& &  {}- O_{ki}^L (g^{a=2}_4-g^{a=2}_3)\Big]\Big\},
\\
\Sigma^{2,\mathrm{E}}_D(W^{+} \tilde{f}^u_L)&=&g^4\sqrt{2}~ {\rm Re} \Big\{
  \Delta(W) \Delta^{*}(\tilde{f}^u_L)~2~
   V_{i1} f^{L*}_{f^uk}\Big[2 O_{ki}^L g^{a=2}_1 \nonumber\\
& &   {}+ O_{ki}^R (g^{a=2}_4-g^{a=2}_3)\Big]\Big\},
\\
\Sigma^{2,\mathrm{E}}_D(\tilde{f}^d_L \tilde{f}^d_L)&=&
 2 g^4|U_{i1}|^2 |f^L_{f^dk}|^2~ |\Delta(\tilde{f}^d_L)|^2 g^{a=2}_2,\\
\Sigma^{2,\mathrm{E}}_D(\tilde{f}^d_L \tilde{f}^u_L) &=& 
  2g^4 {\rm Re}\big\{\Delta(\tilde{f}^d_L)
  \Delta^{*}(\tilde{f}^u_L)  U_{i1}f^{L*}_{f^dk}V_{i1} f^{L*}_{f^uk} (g^{a=2}_4-g^{a=2}_3)\big\},\\
\Sigma^{2,\mathrm{E}}_D(\tilde{f}^u_L \tilde{f}^u_L) &=& 
  -2 g^4|V_{i1}|^2 |f^L_{f^uk}|^2~|\Delta(\tilde{f}^u_L)|^2 g^{a=2}_1,
\end{eqnarray}
where
\begin{eqnarray}
g^{a=2}_1&=&m_i (p_5 p_7) (p_6 s^{a=2}), \label{96}\\
g^{a=2}_2&=&m_i (p_5 p_6) (p_7 s^{a=2}), \\
g^{a=2}_3&=&m_k (p_3 p_7) (p_6 s^{a=2}),\\
g^{a=2}_4&=&m_k (p_3 p_6) (p_7 s^{a=2}).
\end{eqnarray}
The propagators are 
$\Delta(W)=i/((p_3-p_5)^2-m^2_W), 
\Delta(\tilde{f}^u_L)=i/((p_3-p_6)^2-m^2_{\tilde{f}^u_L})$ and 
$\Delta(\tilde{f}^d_L)=i/((p_3-p_7)^2-m^2_{\tilde{f}^d_L})$ and the couplings
are given by
\begin{eqnarray}
f_{f k}^L & = & -\sqrt{2}\bigg[\frac{1}{\cos
  \theta_W}(T_{3f}-e_{f}\sin^2\theta_W)N_{k2}+
  e_{f}\sin \theta_W N_{k1}\bigg],\\
O_{ki}^L & = & -1/\sqrt{2}\Big( \cos\beta N_{k4}-\sin\beta N_{k3}
\Big)V_{i2}^{*}
+\Big( \sin\theta_W N_{k1}+\cos\theta_W N_{k2} \Big)
V_{i1}^{*},\\
O_{ki}^R & = & +1/\sqrt{2}\Big( \sin\beta N^{*}_{k4}+\cos\beta
N^{*}_{k3}\Big) U_{i2}
+\Big( \sin\theta_W N^{*}_{k1}+\cos\theta_W N^{*}_{k2} \Big)
U_{i1},
\end{eqnarray}
where $\tan\beta=v_2/v_1$ is the ratio of the vacuum expectation 
values of the Higgs fields and the unitary $(4\times4)$-matrix $N$ 
diagonalizes the complex symmetric neutralino mass matrix which is 
given in the basis 
$(\tilde{\gamma}, \tilde{Z}, \tilde{H}^0_a, \tilde{H}^0_b)$ \cite{bfm}.

The kinematic functions $g^a_1$, $g^a_2$, $g^a_3$, $g^a_4$, $a=2$  are real.
When multiplied by the purely imaginary $f^{a=2}_5$, Eq.~(\ref{eq_f5_a=2}),
this leads to triple products sensitive to the CP phases 
of the couplings $O^{'L,R}_{ij}$
in the production process,
which in the laboratory system read:
\begin{eqnarray}
g^{a=2}_1\cdot f_5^{a=2} &=& i 2 E_b m_i m_j (p_5 p_7)
\vec{p}_6 ( \vec{p}_1\times \vec{p}_3) , \label{eq_coupl_1}\\
g^{a=2}_2\cdot f_5^{a=2} &=& i 2 E_b m_i m_j (p_5 p_6)
\vec{p}_7 ( \vec{p}_1\times \vec{p}_3) , \label{eq_coupl_2}\\
g^{a=2}_3\cdot f_5^{a=2} &=& i 2 E_b m_j m_k 
(p_3 p_7)\vec{p}_6 (\vec{p}_1 \times \vec{p}_3), \label{eq_coupl_3}\\
g^{a=2}_4\cdot f_5^{a=2} &=& i 2 E_b m_j m_k 
(p_3 p_6)\vec{p}_7 (\vec{p}_1 \times \vec{p}_3).\label{eq_coupl_4}
\end{eqnarray}
As outlined above, these expressions will be
multiplied in Eqs.~(\ref{eq_spgamZ_o})--(\ref{eq_spZsnu_o})
by the factors
$i\cdot {\rm Im}\{(O_{ij}^{'L*}-O_{ij}^{'R*}) \}$
etc., and contribute to the first term of
Eq.~(\ref{CPinT}) and, hence,
to the numerator of the asymmetry
$A_T$, Eq.~(\ref{Asy}).

%%%%%%%%%%%%%%%%%%%%%%%%%%%%%%%%%%%%%%%%%%%%%%%%%%%%%%%%%%
\subsection{T-odd terms of decay and T-even terms of production}
%%%%%%%%%%%%%%%%%%%%%%%%%%%%%%%%%%%%%%%%%%%%%%%%%%%%%%%%%%

The factor
\begin{equation}
\Sigma^{a,\mathrm{O}}_D(\tilde{\chi}^+_i) =
\Sigma^{a,\mathrm{O}}_D(W^+ W^+) +
\Sigma^{a,\mathrm{O}}_D(W^+ \tilde{f}^d_L) +
\Sigma^{a,\mathrm{O}}_D(W^+ \tilde{f}^u_L) +
\Sigma^{a,\mathrm{O}}_D(\tilde{f}^d_L \tilde{f}^u_L)
\end{equation}
in the second term in Eq.~(\ref{CPinT})
with
\begin{eqnarray}
\Sigma^{a,\mathrm{O}}_D(W^{+} W^{+}) &=& 
 2 g^4 |\Delta(W)|^2 
 {\rm Re}\{(O_{ki}^{L*} O_{ki}^R -O_{ki}^L O_{ki}^{R*}) g^a_5\},
  \label{eq_sdWW_o}\\
\Sigma^{a,\mathrm{O}}_D(W^{+} \tilde{f}^d_L) &=& 
  - \sqrt{2}g^4 {\rm Re}\big\{\Delta(W) \Delta^{*}(\tilde{f}^d_L)2
  U^*_{i1} f^L_{f^dk} O_{ki}^L g^a_5 \big\}, \label{eq_sdWsd_o}\\
\Sigma^{a,\mathrm{O}}_D(W^{+} \tilde{f}^u_L) &=& 
  -\sqrt{2}g^4 {\rm Re} \big\{\Delta(W) \Delta^{*}(\tilde{f}^u_L) 2
 V_{i1} f^{L*}_{f^uk}  O_{ki}^R g^a_5\big\},\\
\Sigma^{a,\mathrm{O}}_D(\tilde{f}^d_L \tilde{f}^u_L) &=& 
  - 2 {\rm Re}\big\{\Delta(\tilde{f}^d_L)
  \Delta^{*}(\tilde{f}^u_L) U_{i1}f^{L*}_{f^dk}V_{i1} f^{L*}_{f^uk}
  g^a_5\big\} 
\label{eq_sdsdsu_o}
\end{eqnarray}
is sensitive to CP  violation in the 
decay of the chargino
$\tilde{\chi}^+_i$~\cite{Bartl:2004jj,Moortgat-Pick:1998sk} due to the 
purely imaginary kinematic factor
\begin{equation}
g_5^a=i \cdot m_k \epsilon_{\mu\nu\rho\sigma} s^{a\mu} p_3^{\nu}
p_7^{\rho} p_6^{\sigma} ~. \label{g5-decay}
\end{equation}
For example in Eq.~(\ref{eq_sdWW_o})
it is multiplied by the factor
$i\cdot {\rm Im}\{(O_{ki}^{L*} O_{ki}^R -O_{ki}^L O_{ki}^{R*})\}$,
which depends on the phases $\phi_\mu$ and $\phi_{M_1}$
and contributes to the CP asymmetry $A_{\rm CP}$, Eq.~(\ref{ACP}).
Analogous contributions follow from
Eqs.~(\ref{eq_sdWsd_o})--(\ref{eq_sdsdsu_o}).

The T-even contributions from the production
process in Eq.~(\ref{CPinT}) are
\begin{eqnarray}
\Sigma^{a,\mathrm{E}}_P(\tilde{\chi}^+_i) & = &
\Sigma^{a,\mathrm{E}}_P(\gamma \gamma) +
\Sigma^{a,\mathrm{E}}_P(\gamma Z) +
\Sigma^{a,\mathrm{E}}_P(\gamma \tilde{\nu}) \nonumber\\
&&{}+ \Sigma^{a,\mathrm{E}}_P(Z Z) +
\Sigma^{a,\mathrm{E}}_P(Z \tilde{\nu}) +
\Sigma^{a,\mathrm{E}}_P(\tilde{\nu} \tilde{\nu}), \label{eq_sumdecay_te}
\end{eqnarray}
with
\begin{eqnarray}
\Sigma^{a,\mathrm{E}}_P(\gamma \gamma) &=&
 g^4\sin^2\theta_W |\Delta(\gamma)|^2 c^{P}_{-}(\gamma\gamma)
  \delta_{ij}(-f^a_1+f^a_2+f^a_4-f^a_3),\label{49}\\
\Sigma^{a,\mathrm{E}}_P(\gamma Z) &=& g^4\tan^2\theta_W 
{\rm Re}\Big\{\Delta(\gamma)\Delta^{*}(Z)  \delta_{ij}
  \Big[c^{P}_{+}(\gamma Z)(O_{ij}^{'R*}-O_{ij}^{'L*})
  (f^a_1+f^a_2) \nonumber\\
 & & {}+ c^{P}_{-}(\gamma Z)  (O_{ij}^{'R*}+O_{ij}^{'L*}) (-f^a_1+f^a_2+f^a_4-f^a_3)
      \Big]\Big\},\\
\Sigma^{a,\mathrm{E}}_P(\gamma \tilde{\nu}) &=& -\frac{g^4}{2}\sin^2\theta_W 
{\rm Re}\big\{\Delta(\gamma)
  \Delta^{*}(\tilde{\nu}) \delta_{ij}
  c^{P}_{+}(\gamma\tilde{\nu}) V^*_{i1} V_{j1} (2f^a_2+f^a_4-f^a_3)\big\},\\
\Sigma^{a,\mathrm{E}}_P(ZZ)&=&\frac{g^4}{2\cos^4\theta_W}|\Delta(Z)|^2 
  \Big[c^{P}_{+}(ZZ) (|O_{ij}^{'R}|^2-|O_{ij}^{'L}|^2) (f^a_1+f^a_2)
  \nonumber\\
 & & + c^{P}_{-}(ZZ)\Big((O_{ij}^{'L} O_{ij}^{'R*}+O_{ij}^{'R} O_{ij}^{'L*}) (f^a_4-f^a_3)
  \nonumber\\
 & &     
  +(|O_{ij}^{'R}|^2+|O_{ij}^{'L}|^2) (-f^a_1+f^a_2)
        \Big)\Big], \\
\Sigma^{a,\mathrm{E}}_P(Z \tilde{\nu})&=&
  -\frac{g^4}{2\cos^2\theta_W} {\rm Re}\big\{\Delta(Z) \Delta^{*}(\tilde{\nu}) 
  c^{P}_{+}(Z \tilde{\nu})
  V^*_{i1} V_{j1} \big(2 O_{ij}^{'L} f^a_2 + O_{ij}^{'R}
  (f^a_4-f^a_3)\big)\big\},\nonumber\\
\\
\Sigma^{a,\mathrm{E}}_P(\tilde{\nu} \tilde{\nu})&=&
  -\frac{g^4}{4}|V_{i1}|^2|V_{j1}|^2 |\Delta(\tilde{\nu})|^2  c^{P}_{+}(\tilde{\nu} 
\tilde{\nu}) f^a_2,\label{54}
\end{eqnarray}
where
\begin{eqnarray}
f^a_1&=& m_i (p_2 p_4)(p_1 s^a),\label{44}\\
f^a_2&=& m_i (p_1 p_4)(p_2 s^a),\\
f^a_3&=& m_j (p_2 p_3)(p_1 s^a),\\
f^a_4&=& m_j (p_1 p_3)(p_2 s^a).
\end{eqnarray}
Since $s^{a}(\tilde{\chi}^+_i)$ for $a=2$ is perpendicular to the production plane,
$\Sigma^{2,\mathrm{E}}_P(\tilde{\chi}^+_i)$ 
vanishes, so that in $A_{\rm CP}$ only the contributions of the
longitudinal polarization ($a=3$) and of the transverse polarization in the
production plane ($a=1$)  have to be taken into account.

Finally, the triple products sensitive to the CP phases
in the chargino decay in the laboratory system read
\begin{eqnarray}
\sum_{a=1,3} f^a_1 \cdot g^a_5 &=&
  i m_i m_k (p_2 p_4)\Big\{-E_b \vec{p}_5 (\vec{p}_7\times \vec{p}_6)
  -E_7 \vec{p}_5 (\vec{p}_6\times \vec{p}_1 ) \nonumber\\[-3mm]
 &&\phantom{i m_i m_k (p_2 p_4)\Big\{}
  + E_6 \vec{p}_5 (\vec{p}_7\times \vec{p}_1)
  + E_5 \vec{p}_1 (\vec{p}_7\times \vec{p}_6) \Big\} , \label{eq_te1}\\[2mm]
\sum_{a=1,3} f^a_2 \cdot g^a_5 &=&
  i m_i m_k (p_1 p_4)\Big\{-E_b \vec{p}_5 (\vec{p}_7\times \vec{p}_6)
  +E_7 \vec{p}_5 (\vec{p}_6\times \vec{p}_1 ) \nonumber\\[-3mm]
 &&\phantom{i m_i m_k (p_2 p_4)\Big\{}
  - E_6 \vec{p}_5 (\vec{p}_7\times \vec{p}_1)
  - E_5 \vec{p}_1 (\vec{p}_7\times \vec{p}_6) \Big\} , \label{eq_te2}\\[2mm]
\sum_{a=1,3} f^a_3 \cdot g^a_5 &=&
  i m_j m_k (p_2 p_3)\Big\{-E_b \vec{p}_5 (\vec{p}_7\times \vec{p}_6)
  -E_7 \vec{p}_5 (\vec{p}_6\times \vec{p}_1 ) \nonumber\\[-3mm]
 &&\phantom{i m_i m_k (p_2 p_4)\Big\{}
  + E_6 \vec{p}_5 (\vec{p}_7\times \vec{p}_1)
  + E_5 \vec{p}_1 (\vec{p}_7\times \vec{p}_6) \Big\} , \label{eq_te3}\\[2mm]
\sum_{a=1,3} f^a_4 \cdot g^a_5 &=&
  i m_j m_k (p_1 p_3)\Big\{-E_b \vec{p}_5 (\vec{p}_7\times \vec{p}_6)
  +E_7 \vec{p}_5 (\vec{p}_6\times \vec{p}_1 ) \nonumber\\[-3mm]
 &&\phantom{i m_i m_k (p_2 p_4)\Big\{}
  - E_6 \vec{p}_5 (\vec{p}_7\times \vec{p}_1)
  - E_5 \vec{p}_1 (\vec{p}_7\times \vec{p}_6) \Big\}. \label{eq_te4}
\end{eqnarray}

\end{appendix}

\end{document}